\documentclass[journal=jacsat,manuscript=article]{achemso}

\usepackage[version=3]{mhchem} 
\usepackage{amssymb}
\usepackage{amsmath}
\usepackage{amsthm}
\usepackage{mathtools}
\usepackage[colorlinks,linkcolor=blue, citecolor=red]{hyperref}
\usepackage{mathrsfs}

\author{Weijun Wu}
    \affiliation[Princeton University]{Department of Chemistry, Princeton University, Princeton, New Jersey, 08544, USA}
\author{Gregory D. Scholes}
    \email{gscholes@princeton.edu}
    \affiliation[Princeton University]{Department of Chemistry, Princeton University, Princeton, New Jersey, 08544, USA}

\title[An \textsf{achemso} demo]
  {Foundations of Quantum Information for Physical Chemistry}

\abbreviations{IR,NMR,UV}

\begin{document}

\begin{tocentry}
    \centering\includegraphics[width=\textwidth]{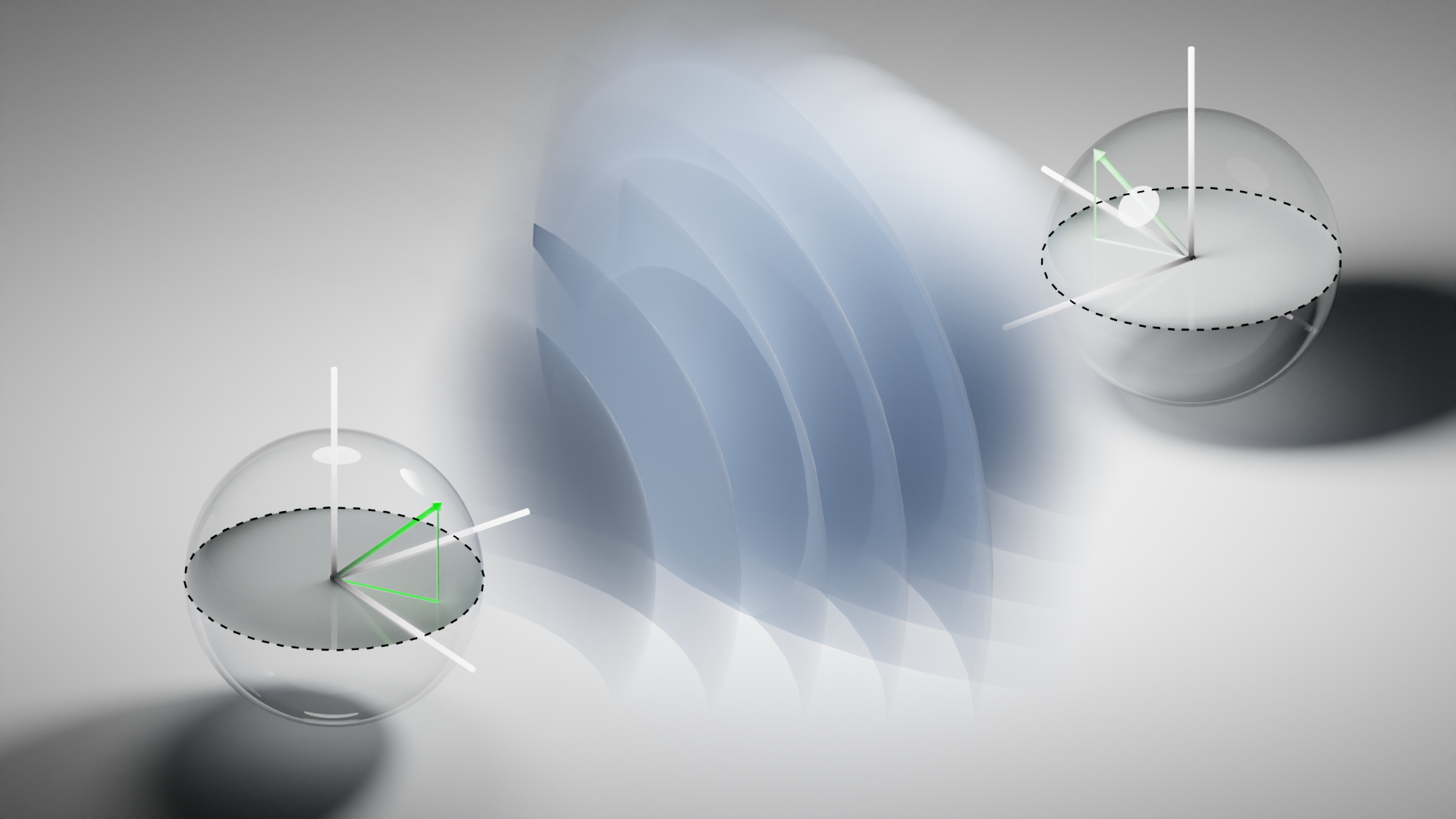}
\end{tocentry}

\begin{abstract}
    Quantum information, a field in which great advances have been made in the past decades, now presents opportunities for chemistry. One roadblock to progress, especially for experimental chemical science, is that new concepts and technical definitions need to be learned. In this paper, we review some basic, but sometimes misunderstood, concepts of quantum information based on the mathematical formulation of quantum mechanics that will be useful for chemists interested in discovering ways that chemistry can contribute to the quantum information field. We cover topics including qubits and their density matrix formalism, quantum measurement as a quantum operation, information theory, and entanglement. We focus on the difference between the concepts in the quantum context and the classic context. We also discuss the relation and distinction among entanglement, correlation, and coherence. We aim to clarify the rigorous definition of these concepts, and then indicate some examples in physical chemistry.
\end{abstract}

\section{Introduction} 
Quantum information, which has become especially prominent in the past decade, demonstrated the potential for new kinds of technology relevant to information processing, communications, and sensors in the framework of quantum mechanics \cite{horodecki2009quantum}. These new ways of handling information are based on exploiting special correlations encoded in quantum states that serve as resources. The field began with the study of nontrivial quantum phenomenon in cold atoms and quantum optics \cite{saffman2010quantum}, and now researchers are also interested in detecting non-trivial quantum phenomena in chemical molecules.\cite{kaufman2021quantum}. Owing to their small size and fundamentally quantum-mechanical nature, molecules may provide interesting building blocks for quantum resources \cite{chitambar2019quantum}. We already know of fundamental examples, particularly based on electron and nuclear spin states. The next challenge is to put quantum correlations to work to produce new kinds of chemical-scale functions.\cite{scholes2023molecular}

With the development of ultrafast spectroscopy that can prepare superpositions and resolve the evolution of superposition states known as wavepackets, coherence phenomena in molecular and biological systems are drawing more attention from the chemistry community. The field opens many fundamental questions such as how to identify functions that arise from underlying nonclassical dynamics. More recently still, there has been interest in working out how chemical systems will be useful for quantum information and related technologies\cite{Wasielewski-review}. To elucidate these opportunities, it is important to appreciate fundamentally what constitutes quantum resources. It is also an open challenge to describe and identify quantum correlations in complex molecular systems. There are many challenges to address at the intersection of chemical molecules and quantum information.  Questions of interest include: How to design molecules that have quantum states serving demonstrably as a resource that enables a quantum function? How to maintain molecular quantum states with long coherence time against a noisy environment? How to think about about unique quantum resources enabled by chemical systems, but not hidden by the inherent complexity of those systems? Ways of addressing these kinds of problems might be inspired by a physical understanding of the basis of quantum information.

Our aim in this paper is to provide a rigorous, but succinct, background of the quantum information field relevant to researchers in physical chemistry. We have tried to use terminology and examples familiar to this target readership. The paper is particularly relevant for spectroscopy and dynamics in physical chemistry, where quantum phenomena such as quantum coherence are of current interest. We will briefly review how to describe quantum states as density matrices, how to control quantum systems via measurement, and how to understand their important features including entanglement, and entropy. We will clarify the relation between the concepts of quantum entanglement, quantum correlation, and quantum coherence. We will focus on the potential advantages of quantum systems that differentiate quantum entanglement and quantum information from classical correlation and classical information. Some examples of chemical systems are provided throughout to illustrate the utility of quantum information principles in a molecular chemistry context. \cite{NielsenChuang, Ballentine, Peres, Barnett}

\section{Qubits and Superposition}

Classical protocols for digital communication and computation are based on bits, like switches that can toggle between 0 and 1 positions. A system of $n$ classical switches can encode $2^n$ different classical states, that is, unique configurations of the switches, e.g. $01110010 \dots$. 

The classical state can be directly generalized to a quantum state, where each switch is replaced by a two-level quantum system, such as a spin-half electron or a single-excitation molecular exciton, known as quantum bits, or qubits. The quantum state of a sequence of $n$ qubits in the orthonormal bases of $|0\rangle$ and $|1\rangle$ can be constructed from the tensor product, like $|0\rangle |1\rangle  |1\rangle |1\rangle |0\rangle |0\rangle |1\rangle |0\rangle \dots$. In total, we obtain $2^n$ unique product states in Hilbert space $\mathcal{H}$, very much like the classical sequence of switches. This set of product states forms one possible basis for $\mathcal{H}$.  

A critical feature that distinguishes the qubits from classical bits is quantum superposition, which is similar to wave interference when added in or out of phase. A qubit may not be on either $|0\rangle$ or $|1\rangle$ but a coherent superposition state \cite{jeong2001quantum}, represented by the wavefunction $|\psi \rangle = c_0|0\rangle + c_1|1\rangle$, where $c_0$ and $c_1$ are complex numbers satisfying the normalization condition $\left| c_0 \right|^2+\left| c_1 \right|^2=1$. Without loss generality, we can take $c_0=\cos \frac{\theta}{2}$, $c_1=\sin \frac{\theta}{2}e^{i\phi}$, where $\theta$ and $\phi$ are in the range $0\leqslant \theta \leqslant \pi$ and $0\leqslant \phi <2\pi$, respectively. We drop the global phase between $|0\rangle$ and $|1\rangle$ because it has no observable effects, so $\phi =\mathrm{Arg}\left( c_1 \right) -\mathrm{Arg}\left( c_0 \right)$ is the relative phase between $|0\rangle$ and $|1\rangle$. The physical meaning of $\theta$ and $\phi$ can be interpreted as the polar angle and the azimuth angle of spherical polar coordinates, which together represent the points on a unit sphere, known as the Bloch sphere. The details of the Bloch sphere are discussed in the next section, as summarized in Figure.~\ref{Fig:BS}. The wavefunction of a qubit can be bijection mapped to a unit vector $\vec{n}$ with the spherical polar direction $\left( \theta ,\phi \right) $ 
\begin{equation}
    |\psi \rangle =|\vec{n}\rangle =\cos \frac{\theta}{2}|0\rangle +\sin \frac{\theta}{2}e^{i\phi}|1\rangle 
\label{equ:qubit.0}
\end{equation}
The wavefunction of a qubit can be expressed by only two real numbers, a unit vector, instead of two complex numbers. 

This wavefunction formalism is essential to account for the phase relationships underlying quantum superposition, or more rigorously, quantum coherence between $|0\rangle$ and $|1\rangle$. Consider the expectation value of an observable $O$ regarding $|\psi \rangle$
\begin{equation}
    \left< O \right> =\langle \psi |O|\psi \rangle =\left| c_0 \right|^2\langle 0|O|0\rangle +\left| c_1 \right|^2\langle 1|O|1\rangle +c_0c_{1}^{*}\langle 1|O|0\rangle +c_1c_{0}^{*}\langle 0|O|1\rangle 
\label{equ:qubit.1}
\end{equation}
The last two terms on RHS of Eq.~\ref{equ:qubit.1} represent the effect of superposition, which might be either constructive or destructive, while in a classical system, we interpret $\left< O \right>$ as the classical probability that our system has property $O$, and the expression is solely comprised of the first two terms in Eq.~\ref{equ:qubit.1}. Thus, superposition is a unique feature in quantum systems. Quantum coherence between $|0\rangle$ and $|1\rangle$ then can be defined by $c_0c_{1}^{*}$ and $c_1c_{0}^{*}$ to quantify the magnitude of superposition. The superposition can be easily generalized to $n$ qubits in $\mathcal{H}$. Some famous examples of the bi-qubit superposition states are the Bell states
\begin{subequations}
\begin{align}
    |\Psi _{\pm}\rangle _{\mathrm{AB}}=\frac{1}{\sqrt{2}}\left( |0\rangle _{\mathrm{A}}\otimes |1\rangle _{\mathrm{B}}\pm|1\rangle _{\mathrm{A}}\otimes |0\rangle _{\mathrm{B}} \right) 
    \\
    |\Phi _{\pm}\rangle _{\mathrm{AB}}=\frac{1}{\sqrt{2}}\left( |0\rangle _{\mathrm{A}}\otimes |0\rangle _{\mathrm{B}}\pm|1\rangle _{\mathrm{A}}\otimes |1\rangle _{\mathrm{B}} \right) 
\end{align} 
\label{equ:qubit.2}
\end{subequations}
where subscripts A and B label the two qubits and AB labels the composite system. Bell states comprise a set of orthonormal bases. Bell states cannot be factored into two local states, signaling entanglement. Non-realism and Non-locality are the non-classical results of superposition\cite{bell1964einstein}, challenged in the Einstein–Podolsky–Rosen (EPR) paradox\cite{EPR}.

The quantum entanglement underlying non-locality and non-realism is the key resource for quantum information. We would like to separate the qubits spatially after entangling them so that local measurement can be done on each qubit. Being able to manipulate qubits in this manner is a particular challenge for molecular-scale systems. A possible example is the entangled pair of electrons and hole in the donor and acceptor separately after exciton dissociation at the interface of organic photovoltaics, but the entanglement lifetime (coherence time) may be short during charge transfer \cite{wu2022polariton}.

\section{Density Matrix Formalism}
Previously we only discussed pure states in wavefunction formalism. Pure states of practical interest provide an excellent foundation for understanding but tend to be fragile in molecular systems. A general way to describe qubits and other quantum systems is density matrix formalism.

\subsection{Mixed State}
The idea of mixed states originates from statistical ensembles, where each `virtual' system represents a possible state $|\psi _i\rangle $ that the real system might be. Statistically, the real system can be on the state $|\psi _i\rangle $ with the probability $p_i$ (satisfying $p_i\geqslant0$ and $\sum_i{p_i}=1$.). The set $\left\{  p_i,|\psi _i\rangle  \right\} $ is known as an ensemble of pure states\cite{NielsenChuang}, or a mixed state. Particularly, in the uniform ensemble $\left\{  1,|\psi \rangle  \right\} $ where all the systems are on the same quantum state, this system can be simply described by the pure state $|\psi \rangle $. 

The ensemble of pure states $\left\{  p_i,|\psi _i\rangle  \right\} $ can be further written as the density matrix
\begin{equation}
    \rho =\sum_i{p_i|\psi _i\rangle \langle \psi _i|}
\label{equ:DM.mixed}
\end{equation}
Eq.~\ref{equ:DM.mixed} shows that the state representing the ensemble will be a statistical mixture of pure states, and a uniform ensemble, i.e. a pure state $|\psi\rangle$ can be similarly expressed as
\begin{equation}
\rho =|\psi \rangle \langle \psi |
\label{equ:DM.pure}
\end{equation}
The redundant global phase is eliminated in the density matrix formalism and only the necessary relative phase in Eq.~\ref{equ:qubit.0} counts. Thus density matrices are the more general and concise expression of quantum states. The expectation value of state $\rho$, that is, the mean expectation value of $\langle \Psi_i |O| \Psi_i \rangle$ for operator $O$ is
\begin{equation}
    \left< O \right> =\sum_i{p_i\langle \Psi _i|O|\Psi _i\rangle}=\mathrm{Tr}\left( O\rho \right) 
    \label{equ:DM.expection}
\end{equation}

Notice that $|\psi _i\rangle$ in Eq.~\ref{equ:DM.pure} do not need to be orthogonal to each other. Thus, Eq.~\ref{equ:DM.mixed} may not be the eigenstructure of $\rho$. One may also diagonalize the density matrix
\begin{equation}
\rho =\sum_{m=1}^N{\lambda _m|m\rangle \langle m|}
\label{equ:DM.eigen}
\end{equation}
where $|m\rangle$ is a set of orthonormal bases in the Hilbert space and  $N=\mathrm{dim}\left( \rho \right) $ is the dimension of this Hilbert space. In this paper, we denote $\lambda _m$ as the eigenvalue of the density matrix and $p_i$ as some probability distribution. One can deduce that the eigenvalues of $\rho$ satisfy $\lambda_m\geqslant0$ and $\sum_{m=1}^N{\lambda _m}=1$. Thus, $\lambda _m$ is also a probability distribution. 

\begin{figure}[!h]
	\centering\includegraphics[width=2.5in]{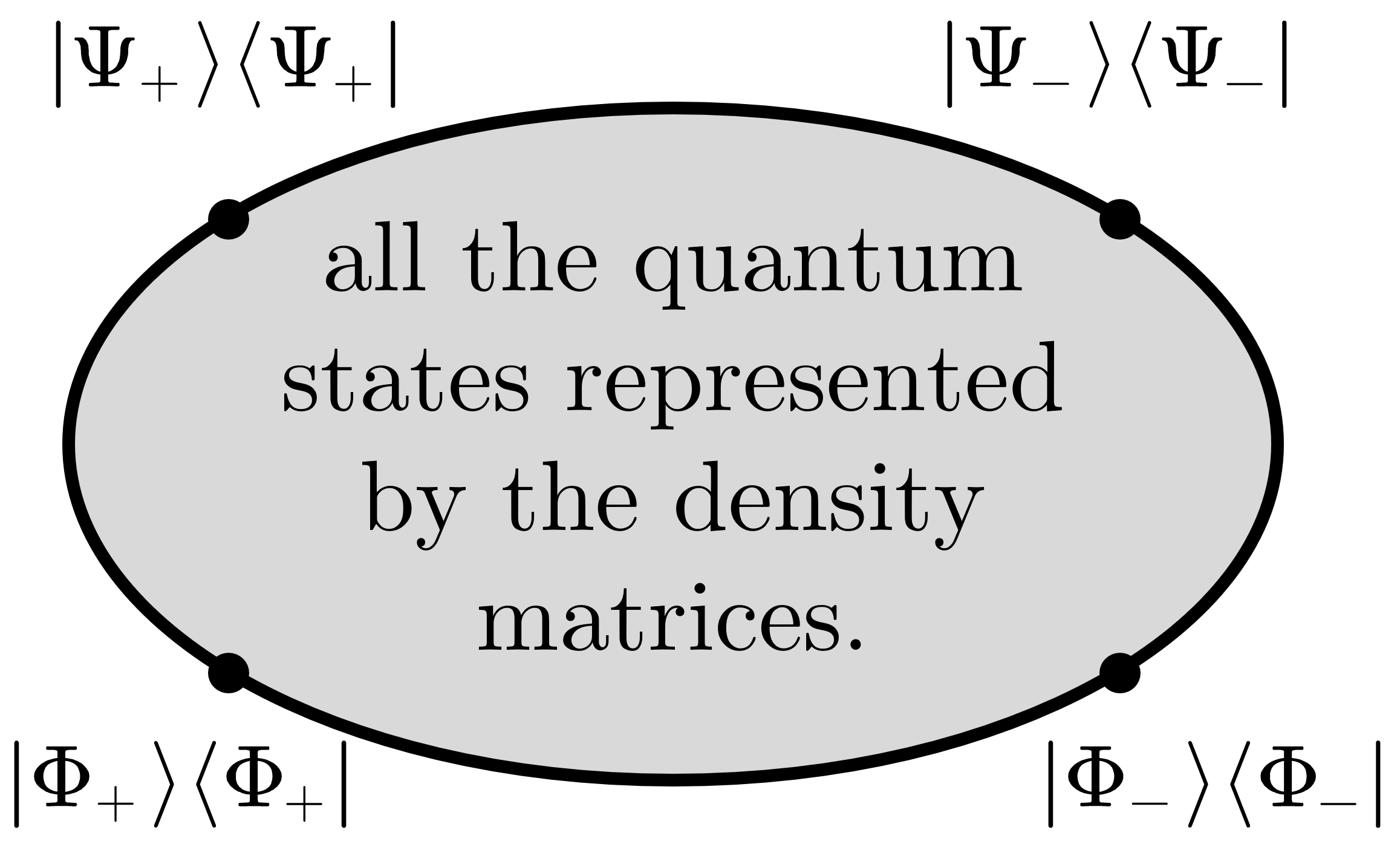}
	\caption{An illustration for convexity of Liouville space. All pure states (the black solid curve) define the convex hull for Liouville space of all the density matrices. These pure states include the Bell states, shown schematically, in the example.}
	\label{Fig:LS}
\end{figure}

The density matrix is Hermitian ($\rho =\rho ^{\dagger}$), positive semi-definite ($\langle \Psi |\rho |\Psi \rangle \geqslant 0$), and unit ($\mathrm{Tr}\left( \rho \right) =1$)\cite{KadRing}. Generally, any operator satisfying these three conditions is a density matrix. The set comprising all the density matrices of a certain dimension is called the Liouville space, which is a space of bounded operators on Hilbert space. Liouville space is a convex set \cite{NielsenChuang}, which means that a convex combination of some density matrices is still a density matrix in Liouville space, i.e. $\rho =\sum_i{p_i\rho _i}$ with $p_i\geqslant 0$ and $\sum_i{p_i}=1$, known as the incoherent superposition. Pure states cannot be a convex combination of other states in Liouville space, so they comprise the convex hull of Liouville space \cite{Webster} (Figure.~\ref{Fig:LS}). This point turns out to be extremely important for quantum information because entanglement (and other quantum correlations) are hidden and diluted in mixed states.

\subsection{Purity}
Even though the definition of density matrix Eq.~\ref{equ:DM.mixed} may not be its eigenstructure, the special case for the pure state (Eq.~\ref{equ:DM.pure}) always is. By comparing Eq.~\ref{equ:DM.pure} and Eq.~\ref{equ:DM.eigen}, the density matrix is pure if and only if it has only one non-zero eigenvalue. This is the qualitative criterion for a pure state. Moreover, we can define the purity of a density matrix to quantify mixedness:
\begin{equation}
\gamma \left( \rho \right) =\mathrm{Tr}\left( \rho ^2 \right)=\left< \rho \right> 
\label{equ:DM.purity}
\end{equation}
Purity is a useful gauge of the mixedness of states in the Liouville space. A more mixed system has a lower purity. Eq.~\ref{equ:DM.purity} means that purity is the expectation value of the density operator. A more uniformly distributed density operator generates a smaller expectation value.

A density matrix $\rho$ is a pure state if and only if $\gamma \left( \rho \right) = 1$, and otherwise $\frac{1}{N}\leqslant \gamma \left( \rho \right)<1$ for mixed states, where $N=\mathrm{dim}\left( \rho \right) $ is the dimension of the $\rho$. To prove this, one can diagonalize the density matrix (Eq.~\ref{equ:DM.eigen}) to calculate $\mathrm{Tr}\left(\rho^2\right) = \sum_{m=1}^N{\lambda _{m}^{2}}$. Because $0\leqslant\lambda_m\leqslant1$, one has $\lambda _{m}^{2}\leqslant \lambda _{m}$. Therefore, $\gamma \left( \rho \right) =\sum_{m=1}^N{\lambda _{m}^{2}}\leqslant \sum_{m=1}^N{\lambda _m}=1$. The equality condition reaches when $\lambda _{m}^{2}= \lambda _{m}$, i.e. $\lambda _m=0,1$. Consider the unit condition $\sum_{m=1}^N{\lambda _m}=1$, only the pure state has purity of 1. The lower bound of $\gamma \left( \rho \right)$ can be proved by Cauchy–Schwarz inequality: $\left( \sum_{m=1}^N{\lambda _m\times 1} \right) ^2\leqslant \left( \sum_{m=1}^N{\lambda _{m}^{2}} \right) \left( \sum_{m=1}^N{1^2} \right) $, which ends up with $\frac{1}{N}\leqslant \gamma \left( \rho \right)$. The equality condition for the lower bound is $\lambda_m = \frac{1}{N}$, i.e. $\rho =\frac{1}{N}I$. The normalized identical operator refers to a completely mixed state, where each eigenstate contributes the same probability.

Purity has many applications in chemistry. For example, in a molecular aggregate system, a widely-used tool to describe the molecular exciton delocalization \cite{kramer1993localization} is inverse participation ratio (IPR) \cite{evers2000fluctuations, wegner1980inverse}, defined as the second moment of the probability density:
\begin{equation}
    \mathrm{IPR}=\sum_x{\left| \langle x|\psi \rangle \right|^4}
\label{equ:DM.ipr}
\end{equation}
where $|\psi \rangle$ is the exciton state and $x$ labels the site of a localized exciton. An exciton state with a smaller IPR implies more delocalization. IPR is the purity of the state after projection-valued measurement (PVM, see Eq.~\ref{equ:Measurement.PVM}), $\mathrm{IPR}=\gamma \left( \rho^{\prime}\right)$, which means delocalization can be described by post-measurement purity.

\subsection{Bloch Sphere}
\begin{figure}[!h]
	\centering\includegraphics[width=1.0\textwidth]{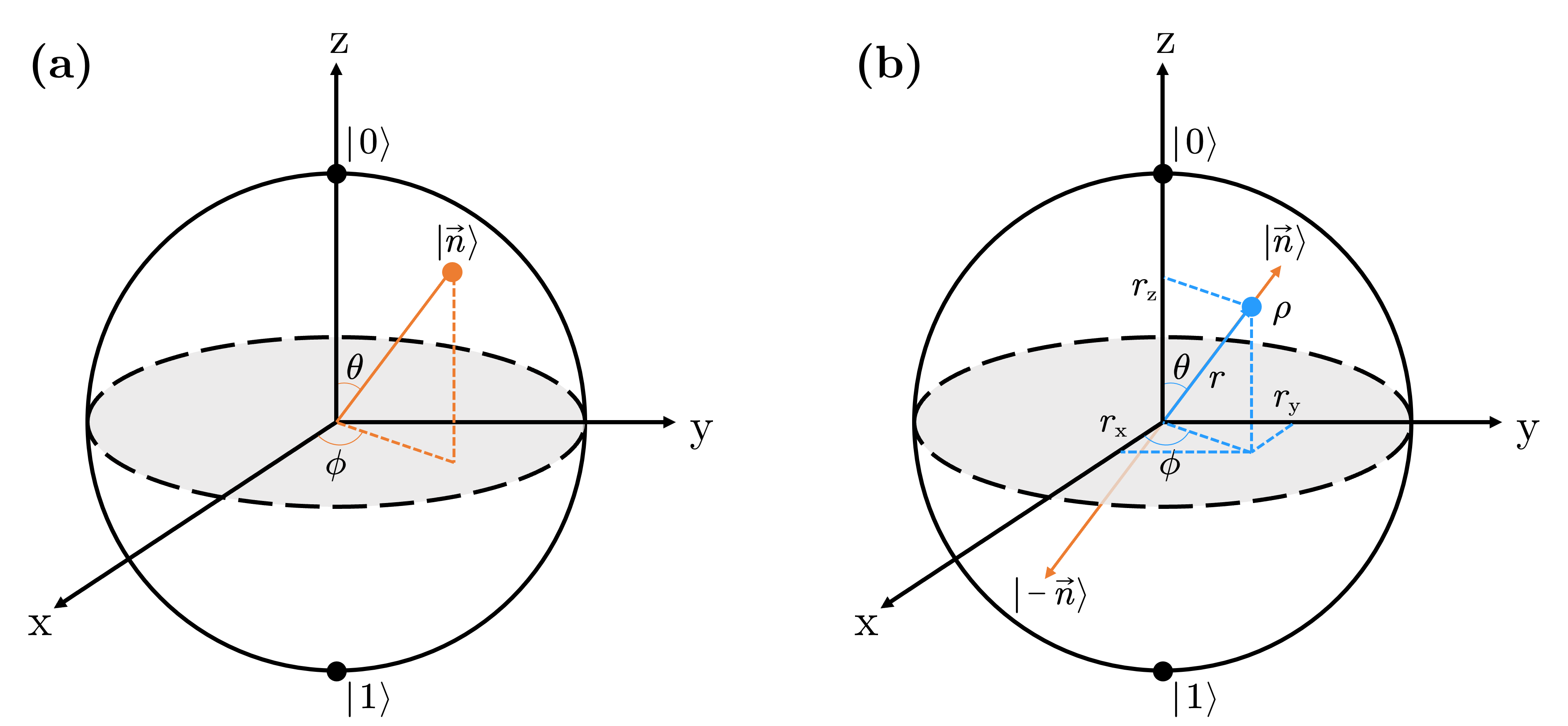}
	\caption{An illustration for Bloch Sphere. (a) Any pure state $|\psi \rangle$ (orange) on Bloch Sphere can be represented by a unit vector $\vec{n}$, in $\left( \theta ,\phi \right) $ coordinates. (b) Any mixed state $\rho$ (blue) inside Bloch sphere can be represented by a Bloch vector $\vec{r}$ in $\left( r_{\mathrm{x}},r_{\mathrm{y}},r_{\mathrm{z}} \right)$ or $\left( r,\theta,\phi \right)$ coordinates, whose eigenvectors are $|\vec{n}\rangle$ and $|-\vec{n}\rangle$.}
	\label{Fig:BS}
\end{figure}

Now we can generalize Eq.~\ref{equ:qubit.0} to describe a single qubit beyond the pure state in the density matrix formalism, i.e. a two-dimensional density matrix $\rho$. It is easy to prove that it can always be decomposed as
\begin{equation}
\rho =\frac{1}{2}\left( I+\vec{r}\cdot \vec{\sigma} \right) =\frac{1}{2}\left( I+r_{\mathrm{x}}\sigma _{\mathrm{x}}+r_{\mathrm{y}}\sigma _{\mathrm{y}}+r_{\mathrm{z}}\sigma _{\mathrm{z}} \right) 
\label{equ:DM.BSqubit}
\end{equation}
where $\vec{\sigma}=\left( \sigma _{\mathrm{x}},\sigma _{\mathrm{y}},\sigma _{\mathrm{z}} \right)$ is the Pauli vector with the three components being the Pauli matrices.  $\vec{r}$  is known as the Bloch vector. The length of Bloch vector $r=\left| \vec{r} \right|$ is in the range $0\leqslant r \leqslant 1$ to satisfy the positive semi-definite condition. The Bloch vector can be explicitly calculated by
\begin{equation}
\vec{r}=\left( r_{\mathrm{x}},r_{\mathrm{y}},r_{\mathrm{z}} \right) =\mathrm{Tr}\left( \vec{\sigma}\rho \right) 
\label{equ:DM.BlochVector}
\end{equation}
where each component is $r_{\alpha}=\mathrm{Tr}\left( \sigma _{\alpha}\rho \right)$,  owing to the relation that $\mathrm{Tr}\left( \sigma _{\alpha}\sigma _{\beta} \right) =2\delta _{\alpha ,\beta}$, $\alpha, \beta=\mathrm{x},\mathrm{y},\mathrm{z}$. Generally, the Pauli matrices together with the identity operator, $\left\{ I,\sigma _{\mathrm{x}},\sigma _{\mathrm{y}},\sigma _{\mathrm{z}} \right\} $,  form a set of complete orthonormal bases spanning the single-qubit Liouville space regarding the inner product defined as $\left< \rho _1,\rho _2 \right> =\frac{1}{2}\mathrm{Tr}\left( {\rho _1}^{\dagger}\rho _2 \right)$. Thus, there exists a one-to-one correspondence between a single-qubit state and a Bloch vector (Eq~.\ref{equ:DM.BSqubit}), i.e. a bijection mapping from $\rho$ to $\vec{r}$. Considering $0\leqslant r \leqslant 1$, Bloch vectors are the points inside or on the surface of a unit sphere, the Bloch sphere, as shown in Figure.~\ref{Fig:BS}.

The vector length $r$ of the Bloch vector is an indicator for the mixedness of single-qubit states. To see this, one can calculate the purity of $\rho$
\begin{equation}
\gamma \left( \rho \right) =\frac{1+r^2}{2}
\label{equ:DM.BSPurity}
\end{equation}
When $r=1$, $\gamma \left( \rho \right)=1$, $\rho$ is a pure state. When $0\leqslant r < 1$, $\frac{1}{2}\leqslant \gamma \left( \rho \right) < 1$, $\rho$ is a mixed state. When $r=0$, $\gamma \left( \rho \right)=\frac{1}{2}$, $\rho$ is a completely mixed state. Thus, Bloch vectors on the Bloch sphere represent the pure states, while inside the Bloch sphere, they represent the mixed states. The closer a state is to the center of the Bloch sphere, the more mixed it can be.

It is also convenient to transform Bloch vectors from Cartesian coordinates to a spherical coordinate system $\vec{r}=\left( r_{\mathrm{x}},r_{\mathrm{y}},r_{\mathrm{z}} \right) =\left( r,\theta,\phi \right)$, where $r_{\mathrm{x}}=r\sin \theta \cos \phi $, $ r_{\mathrm{y}}=r\sin \theta \sin \phi $, $r_{\mathrm{z}}=r\cos \theta $. In the special case of a pure state ($r=1$), we can find the corresponding wavefunction $\rho = \frac{1}{2}\left( I+\vec{n}\cdot \vec{\sigma} \right) =|\vec{n}\rangle \langle \vec{n}|$, where $\vec{n}=\vec{r}$ and $|\vec{n}\rangle$ is defined as Eq.~\ref{equ:qubit.0}. The polar angle and azimuthal angle in Eq.~\ref{equ:qubit.0} are the same as $\left( \theta ,\phi \right) $ in this section, so we use consistent notation. The wavefunction formalism for a single qubit can be recovered from the density matrix formalism with a Bloch sphere. Generally, given any Bloch vector in the spherical coordinate system, one can diagonalize $\rho$
\begin{equation}
\rho =\frac{1+r}{2}|\vec{n}\rangle \langle \vec{n}|+\frac{1-r}{2}|-\vec{n}\rangle \langle -\vec{n}|
\label{equ:DM.BSEigenSpectrum}
\end{equation}
where $\vec{n}=\frac{\vec{r}}{r}$ is the unit vector with the same direction $\left( \theta ,\phi \right) $ as the Bloch vector $\vec{r}$, and $-\vec{n}$ is of the opposite direction. $|\vec{n}\rangle$ and $|-\vec{n}\rangle$ are orthogonal to each other,that is, $\langle \vec{n}|-\vec{n}\rangle =0$. 

The Bloch sphere provides a useful means of visualizing the state of a single qubit and often serves as an excellent test bed for ideas about quantum information \cite{NielsenChuang}. The quantum dynamics of a single qubit maps to the dynamics of a Bloch vector. For example, unitary evolution fixes the Bloch vector on a sphere because $r$ is invariant. In the orthonormal basis of $|0\rangle$ and $|1\rangle$, dephasing means the decrease of $r_{\mathrm{x}}$ and $r_{\mathrm{y}}$, while decay from $|1\rangle$ to $|0\rangle$ means increase of $r_{\mathrm{z}}$. The phase-space formulation for spins can be visualized in the Bloch sphere. For multiple qubits, if we are only interested in the coherent spin state or spin squeezing \cite{ma2011quantum, kitagawa1993squeezed}, the Bloch sphere is also a strong tool. However, a general description for $N$-qubits is limited because the general Bloch sphere lies in a $\left(4^N-1\right)$-dimension space.

\section{Quantum Measurement}
In experiments, we perform measurements by projecting a state onto our measuring apparatus and getting a sequence of outcomes. In the case of a classical system, our ensemble yields properties of the probability distribution with respect to an appropriate random variable. It is reasonable to believe that the measurement apparatus only interacts with the system with a negligible perturbation and therefore does not irreversibly change the observed system. This feature for classical systems is called realism \cite{EPR}.

The challenge for measurements on quantum systems is to construct an analogous map from the set of states to a probability distribution\cite{Holevo} such that we account for the quantum phase encoded in superposition and quantum coherence. A closed quantum system, which ideally does not interact with the rest of the world, evolves according to unitary evolution, without losing quantum coherence. However, when the measuring apparatus, an external system, observes the system to obtain the information inside the system, the interaction makes the system open. Thus, measurement involves a non-unitary evolution of the system, which may destroy the fragile quantum coherence. One may think about studying a large and isolated composite system, the observed system together with the external system, which again obeys unitary evolution. But in practice, it is hard to handle such a closed system because usually we know little about the environment and a large system means an exponential increase in mathematical complexity. So we need other ways to describe the system locally when measuring it, which is called Quantum Measurement Theory. Quantum Measurement Theory is sometimes considered one of the fundamental postulates of quantum mechanics \cite{NielsenChuang}.

\subsection{Projective Measurements and PVM}
Given a system prepared initially at a pure state $|\psi \rangle $, one can carry out the projective measurement with a series of projection operators $\left\{ P_m \right\} $, which are Hermitian ($P_{m}^{\dagger}=P_m$), orthogonal ($P_mP_{m^{\prime}}=\delta _{m,m^{\prime}}P_m$), and complete ($\sum_m{P_m}=I$). The measurement operators satisfying these three conditions are also called Projection-Valued Measure (PVM). Projective measurement aims to project $|\psi \rangle $ from Hilbert space $\mathcal{H}$ to the subspaces $\mathcal{H} _m$. Completeness insures $\mathcal{H} =\bigoplus_m{\mathcal{H} _m}$. After the measurement, the probability that the result $m$ occurs is $p_m=\langle \psi |P_m|\psi \rangle $ and the corresponding quantum state is $\frac{P_m|\psi \rangle}{\sqrt{\langle \psi |P_m|\psi \rangle}}$. A mixed state is generated from a pure state by PVM with the density matrix
\begin{equation}
|\psi \rangle \xrightarrow{\mathrm{PVM}}\rho^{\prime}=\sum_m{P_m|\psi \rangle \langle \psi |P_m}
\label{equ:Measurement.PVM}
\end{equation}

In some cases, it is useful to describe PVM by an observable, $M$. $P_m$ is chosen to be the spectral decomposition: $M=\sum_m{\epsilon _mP_m}$. The spectral theorem \cite{lengyel1936elementary} guarantees Hermiticity, orthogonality, and completeness. To understand this, we can recall the wavefunction collapse theory. When we measure the system with $M$ (in the eigenbasis $|m\rangle $ of $M$), the wavefunction $|\psi \rangle $ will randomly collapse to a certain $|m\rangle $ with probability $\left| \langle m|\psi \rangle \right|^2$ and observation value $\epsilon _m$. From the view of the ensemble, after the measurement we obtain a density matrix $\rho ^{\prime}=\sum_m{\left| \langle m|\psi \rangle \right|^2|m\rangle \langle m|}$, and the expectation value of $M$ is $\left< M \right> =\sum_m{\left| \langle m|\psi \rangle \right|^2\epsilon _m}$. This is a process of PVM where we take $P_m=|m\rangle \langle m|$. Operator $M$ is so special because PVM does not change the expectation value of $M$, since $\langle \psi |M|\psi \rangle =\mathrm{Tr}\left( M\rho ^{\prime} \right) $.

PVM is very useful because most physical systems can only be measured in a very coarse manner. PVM has a property known as repeatability\cite{NielsenChuang}, which means if we perform a PVM once and obtain the outcome $m$, repeating the same PVM gives the outcome $m$ again and does not change the state. Repeatability originates from the orthogonality of PVM operators, which is not guaranteed for many other quantum measurements. Therefore, general measurement theory is extremely useful when discussing quantum measurements.

\subsection{General Measurement and POVM}
General Measurement provides a general perspective to construct a density matrix $\rho^{\prime}$. Similar to the repeated test in the laboratory, by preparing infinity copies of the initial state $|\psi \rangle $ and doing the same measurement to each copy, we can summarize all the possible outcomes (labeled by $m$) and the corresponding probability $p_m$. We then have $\rho^{\prime} =\sum_m{p_m|\psi _m\rangle \langle \psi _m|}$, which matches the definition of density matrix Eq.~\ref{equ:DM.mixed}. So we say General Measurement is an equivalent way to define density matrices, i.e. the ensemble of post-measurement. This procedure works intuitively because the structure of the density matrices reflects the probability distribution of the states. Sometimes the measurement may probe a specific spectral window or interval, so then we only include those states whose expectation values lie in that window and keep other states with zero probability. Same as Eq.~\ref{equ:DM.mixed}, outcome states $|\psi _m\rangle$ do not have to be orthogonal to each other. This feature distinguishes General Measurement from projective measurement.

Rigorously, General Measurement is described by a set of measurement operators $\left\{ M_m \right\}$ that satisfy completeness $\sum_m{M_{m}^{\dagger}M_m}=I$. These are operators, which do not have to be Hermitian, on the state space of the system being measured. Each outcome $m$ (also known as quantum channel $m$) occurs with a probability 
\begin{equation}
p_m=\langle \psi |M_{m}^{\dagger}M_m|\psi \rangle 
\label{equ:Measurement.GMp}
\end{equation}
and results in the quantum state
\begin{equation}
|\psi _m\rangle =\frac{M_m|\psi \rangle}{\sqrt{\langle \psi |M_{m}^{\dagger}M_m|\psi \rangle}}
\label{equ:Measurement.GMs}
\end{equation}
The density matrix after general measurement can be written as
\begin{equation}
|\psi \rangle \xrightarrow{\mathrm{POVM}}\rho ^{\prime}=\sum_m{M_m|\psi \rangle \langle \psi |M_{m}^{\dagger}}
\label{equ:Measurement.GMDM}
\end{equation}

If additional conditions of Hermiticity ($M_{m}^{\dagger}=M_m$) and orthogonality ($M_m^{\dagger}M_{m^{\prime}}=\delta _{m,m^{\prime}}M_m$) are imposed on general measurement, we recover PVM. So PVM is a special case of general measurement. Moreover, PVM on a composite system augmented by unitary operations turns out to be completely equivalent to general measurements on subsystems \cite{NielsenChuang}. 

Sometimes, the post-measurement state of the system Eq~.\ref{equ:Measurement.GMs} is of little interest, and we care more about the probabilities of the respective measurement outcomes, because many statistical properties only depend on the probabilities. In this case, we can rewrite general measurement as $F_m=M_{m}^{\dagger}M_m$, so that the probability is 
\begin{equation}
p_m=\langle \psi |F_m|\psi \rangle 
\label{equ:Measurement.POVMp}
\end{equation}
The set of operators $\left\{ F_m \right\} $ is known as Positive Operator-Valued Measure (POVM), which is sufficient to determine the probabilities of the different measurement outcomes. To avoid any confusion, in this paper, `positive' means `positive semi-definite'. Rigorously, the definition of POVM is a set of operators $\left\{ F_m \right\} $ that is positive semi-definite ($\langle \phi |F_m|\phi \rangle \geqslant 0$) and complete ($\sum_m{F_m}=I$). POVM is simple and intuitive when only the measurement statistics matter. If the post-measurement states are needed, one can always decompose $F_m$ to recover $M_m$ by $F_m=M_{m}^{\dagger}M_m$. PVM is also a special case of POVM where $F_m=M_m$.

If the initial state is prepared in a mixed state, Eq.~\ref{equ:Measurement.GMp}, Eq.~\ref{equ:Measurement.GMs} and Eq.~\ref{equ:Measurement.GMDM} can be generalized to

\begin{equation}
    p_m=\mathrm{Tr}\left( M_m\rho M_{m}^{\dagger} \right) =\mathrm{Tr}\left( \rho F_m \right) 
\label{equ:Measurement.DMGMp}
\end{equation}

\begin{equation}
    \rho _m=\frac{M_m\rho M_{m}^{\dagger}}{\mathrm{Tr}\left( M_m\rho M_{m}^{\dagger} \right)}
\label{equ:Measurement.DMGMs}
\end{equation}

\begin{equation}
    \rho \xrightarrow{\mathrm{POVM}}\rho ^{\prime}=\sum_m{p_m\rho _m}=\sum_m{M_m\rho M_{m}^{\dagger}}
\label{equ:Measurement.DMGMDM}
\end{equation}
In quantum operation, Eq.~\ref{equ:Measurement.DMGMDM} is known as Kraus representation \cite{tong2006kraus}.

Measurement is an important part of quantum information\cite{Holevo, Kraus, QMeasurement}.  In addition to describing measurements of discrete quantum variables (systems with discrete spectra)\cite{davies1970operational}, PVM and POVM can be extended to account for continuous measurement\cite{Ozawa1984, JacobsSteck}. A theory for continuous measurement is needed when we must account for the time it takes to complete the measuring process.

\section{Information Theory and Entropy}
In this section, we will discuss how to understand the information in a quantum system, compared to an analogous classical system. For example, $n$ classical bits can encode $2^n$ distinct binary numbers, which is the same as the dimension of the Hilbert Space for $n$ qubits. The maximum information encoded in either case is the same. However, the quantum advantage for conveying information comes from using compression strategies that are only available to quantum systems. To start, we need to define the amount of information\cite{Stone, Ash}, and we do so using entropy.

\subsection{Classical Information and Shannon Entropy}
Quantification of classical information was devised by Claude Shannon \cite{shannon1948mathematical} as an analogy to Gibbs entropy, known as Shannon entropy. Given a discrete random variable $X$ distributed according to the probability mass function $p(x)$, the Shannon entropy is defined as
\begin{equation}
    H(X) = H(\left\{ p(x) \right\} ) = - \sum_x p(x) \ln{p(x)}
    \label{equ:Entropy.Shannon}
\end{equation}
Intuitively, the amount of information we obtain from the occurrence of a certain event $X$ is $-\ln p\left( X \right)$. The total Shannon entropy can then be understood as the average amount of information for measuring this random variable: $H(X)=E\left[ -\ln p\left( X \right) \right] $. Shannon entropy only depends on the probability distribution. Shannon entropy is defined in this way to ensure that the total information learned from independent events is the sum of the information learned from each event:
\begin{equation}
    H(\left\{ p_{X,Y}(x,y)=p_X(x)p_Y(y) \right\} )=H(\left\{ p_X(x) \right\} )+H(\left\{ p_Y(y) \right\} )
    \label{equ:Entropy.accumulation}
\end{equation}

We can view the entropy either as a measure of our uncertainty before we learn the value of $X$, or as a measure of how much information we have gained after we learn the value of $X$ \cite{NielsenChuang}. To understand this, we can ask a question: when transmitting information in a message (e.g. by code, telephone, etc), how much error---or uncertainty---can be tolerated? For exxmple [sic], cxn I get my messxge xcross sufficiently clexrly if every `a' is erroneously trxnsmitted as `x'? Obviously, this error is not catastrophic because we use the information provided by the other letters together with our knowledge about words in the English language, and possibly other information gathered by experience, to interpret the message correctly. We can explain this concept more rigorously with the following scenario.

In the next examples, we are going to think about searching for particular pairs of letters on a random page of an English text. Suppose that we are interested only in two letters in our document, `Q' and `T'. We scan through the words until we find either `Q' or `T'. The question is, how much information does identifying the letter give us? To answer that, we can normalize the probabilities for the events of finding `Q' and finding `T' as $p\left( \mathrm{Q} \right) =0.01$ and $p\left( \mathrm{T} \right) =0.99$. Because $p\left( \mathrm{T} \right) \gg p\left( \mathrm{Q} \right) $, the first `Q' or `T' we find will almost certainly be `T'. Thus, one measurement resulting in `T' provides very little information because it is unlikely to change our expectations, where the amount of information we gain from this measurement is $-\ln p\left( \mathrm{T} \right) =0.010$. However, one measurement resulting in `Q' provides a lot of information, because a rare event occurs beyond our expectation, where the amount of information we gain from this measurement is $-\ln p\left( \mathrm{Q} \right) =4.605$. From a statistical perspective, the average information we gain from the measurement is $H\left( \left\{ p\left( \mathrm{Q} \right),p\left( \mathrm{T} \right) \right\} \right)\approx 0.056$. The Shannon entropy in this example is quite small because, before any measurement, we anticipate getting `T'. In contrast, if we are interested in the letters `M' and `W', the probability distribution for finding `M' and finding `W' will be $p\left( \mathrm{M} \right) =0.5$ and $p\left( \mathrm{W} \right) =0.5$. Now the measurement provides the most possible information because we really cannot guess whether the first `M' or `W' on the page will be `M' or `W'. The Shannon entropy is as large as $H\left( \left\{ p\left( \mathrm{W} \right) ,p\left( \mathrm{M} \right) \right\} \right)\approx 0.69$. This example also demonstrates that Shannon entropy describes the divergence of a series of probabilities.

\subsection{Quantum Information and Von Neumann entropy}
Classical probability has its analogy in quantum mechanics as a density matrix, where coherence is considered. Quantum information\cite{Barnett} is quantified with respect to the density matrix of the quantum system via the von Neumann entropy:
\begin{equation}
    S\left(\rho\right) = - \operatorname{Tr}\left(\rho \ln{\rho} \right) = \left< -\ln \rho \right> 
    \label{equ:Entropy.VonNeumann}
\end{equation}
Here we naturally replace the classical probability distribution in Eq.~\ref{equ:Entropy.Shannon} by the density matrix. In statistical physics, taking $\rho$ in Eq.~\ref{equ:Entropy.VonNeumann} as a thermal ensemble, $\rho =\frac{1}{Z}e^{-H/k_{\mathrm{B}}T}$, recovers Gibbs entropy. To calculate $S\left(\rho\right)$ practically, one may diagonalize $\rho$ by Eq.~\ref{equ:DM.eigen} and have $S\left( \rho \right) =-\sum_{m=1}^N{\lambda _m\ln \lambda _m}$.

Different from the classical system, each measurement we do to the system will lead to a non-unitary evolution. So the post-measurement will not perfectly recover the information carried by the initial density matrix. Suppose a PVM $\left\{ P_m \right\} $ is performed on a quantum system $\rho$, but we never learn the result of the measurement. It can be proven that \cite{NielsenChuang}
\begin{equation}
    S\left( \rho^{\prime} \right) = S\left( \sum_m{P_m\rho P_m} \right)  \geqslant S\left( \rho \right) 
    \label{equ:Entropy.PVM}
\end{equation}
where the post-measurement density matrix $\rho^{\prime}$ is defined by Eq~.\ref{equ:Measurement.PVM}. The equality condition is reached if and only if $\rho^{\prime} = \rho$, which means $\rho$ is orthogonal in the basis of $\left\{ P_m \right\} $. Notice that PVM will destroy all the coherence on the basis of $\left\{ P_m \right\} $ and make $\rho^{\prime}$ classical-like. So $S\left( \rho^{\prime} \right)$ is the same as Shannon entropy $H(\left\{ \mathrm{Tr}\left( \rho P_m \right) \right\} )$. This relation underlies an advantage for quantum information, whereby the information transfer capacity of a quantum system is constrained differently than that of a classical system\cite{YuenOzawa}.

While one qubit does not convey more information than one classical bit, the quantum correlations in an entangled system can be exploited to enable information to be more compressed. The idea is along the lines of superdense coding \cite{NielsenChuang}. A notable advantage of quantum information, therefore,  is that quantum dense coding allows quantum communication channels to have greater capacity than comparable classical channels\cite{DenseCoding, QCommunication}. A second advantage is that quantum mechanical laws can be leveraged to make communication more secure. Communications can be encrypted in new ways such that it is difficult for an eavesdropper to copy information, and quantum `keys' can be more difficult to crack\cite{QuantumCrypt1, QuantumCrypt2, Barnett}.

\section{Quantum Entanglement}
Entanglement is a central concept in quantum information. In this section, we will give a rigorous definition of quantum entanglement. However, it is difficult (NP-hard) to pinpoint a general state (arbitrary number of particles and arbitrary dimension) as being entangled \cite{gurvits2004classical}, and thus difficult to determine the amount of entanglement \cite{VedralPlenioKnight}. Luckily, there is a standard entanglement measure for pure bipartite states and there are several non-standard entanglement measures for mixed two-qubit systems \cite{simon2000peres}. For more general cases, entanglement witnesses offer some sufficient conditions for entanglement.

\subsection{Separability}
\begin{figure}[!h]
    \centering\includegraphics[width=0.9\textwidth]{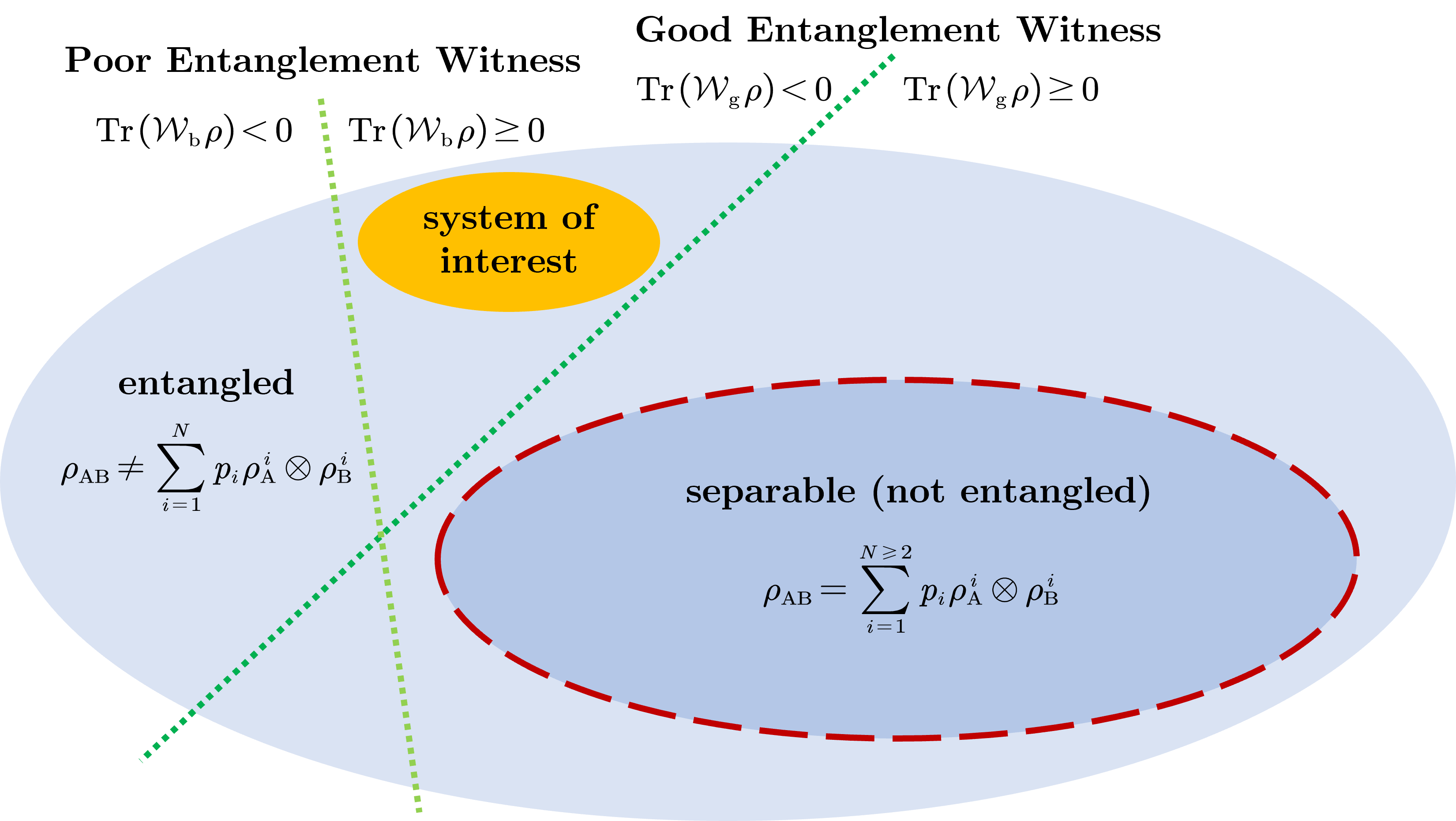}
    \caption{A set of all the bipartite states (Liouville space) as an illustration for entanglement and witness. The red dashed curve is the boundary between entangled states and separable states. The subset of all the separable states is convex. The two green dotted lines represent two examples of hyperplanes that work as entanglement witnesses $\mathcal{W}$. However, only $\mathcal{W}_{\mathrm{g}}$ is a good entanglement witness with respect to the system of interest.}
    \label{fig:EW}
\end{figure}

In quantum mechanics, the opposite of entanglement is separability \cite{Werner1989}, which has features corresponding to classical correlation or even uncorrelatedness (See Figure.~\ref{fig:Liouville}). Separable states are the quantum states of a composite system that can be factorized into individual states belonging to separate subsystems. For example, a bipartite pure state is separable if it can be written as a product state
\begin{equation}
|\Psi \rangle =|\psi \rangle _{\mathrm{A}}\otimes |\psi \rangle _{\mathrm{B}}
\label{equ:4.4}
\end{equation}
More generally, a bipartite mixed state is separable if it can decompose into a sum of tensor products
\begin{equation}
\rho _{\mathrm{AB}}=\sum_i{p_i\rho _{\mathrm{A}}^{i}\otimes \rho _{\mathrm{B}}^{i}}
\label{equ:4.5}
\end{equation}
where $p_i>0$ and $\sum_i{p_i}=1$. Without loss of generality, the separable mixed state can be equivalently defined based on separable pure states
\begin{equation}
\rho _{\mathrm{AB}}=\sum_i{q_i|\Psi ^i\rangle _{\mathrm{AB}}\langle \Psi ^i|_{\mathrm{AB}}}
\label{equ:4.6}
\end{equation}
where $q_i>0$ and $\sum_i{q_i}=1$, and $|\Psi ^i\rangle _{\mathrm{AB}}$ are series of separable pure states. Eq.~\ref{equ:4.5} keeps the convexity of separable density matrices (any convex combination of separable density matrices is separable, as shown in Figure.~\ref{fig:EW}), and Eq.~\ref{equ:4.6} shows that the pure product states (Eq.~\ref{equ:4.4}) comprise the convex hull of the separable set. The definition of separable states (Eq.~\ref{equ:4.5}) can be easily generalized to multipartite states by the tensor product of more subsystems.

Then we can give a rigorous mathematical definition of entanglement. A non-separable quantum state is called an entangled state, i.e. entangled states cannot be convexly decomposed as tensor product states: 
\begin{equation}
\rho _{\mathrm{AB}}\ne\sum_i{p_i\rho _{\mathrm{A}}^{i}\otimes \rho _{\mathrm{B}}^{i}}
\label{equ:4.66}
\end{equation}
The distinction between entanglement and separability is illustrated in Figure.~\ref{fig:EW} by the dashed red curve. The convexity property does not hold for entangled states (Figure.~\ref{fig:EW}). 

Separable states are important in quantum chemistry because they have classical and intuitive analogies. As an example of separable pure bipartite states, the Born-Oppenheimer approximation manually divides the molecules into a bipartite system (the electron subsystem and the nucleus subsystem), further assuming that the total wavefunction is a product state of the electron wavefunction and the nucleus wavefunction. This is a separable state, for which we can study electronic dynamics and nuclear dynamics separately. However, the breakdown of the Born-Oppenheimer approximation will lead to an entangled molecular state, which requires a more sophisticated treatment like non-adiabatic dynamics \cite{tully2012perspective}. Another example is Markov approximation in the open quantum system approach, which is valid when the bath is disentangled from the system so that the bath loses its memory during the dynamics, such as Lindblad or Redfield dynamics \cite{wu2022polariton}. However, the breakdown of the Markov approximation requires the non-Markovian dynamics \cite{breuer2016colloquium}.

\subsection{Schmidt Decomposition and Entanglement Entropy}

Although the criterion of separability and quantifying entanglement of a general multipartite system or mixed state is an NP-hard problem \cite{gurvits2004classical}, there is a rigorous and standard approach to quantifying entanglement of a bipartite pure state, based on Schmidt decomposition.

Schmidt decomposition \cite{acin2000generalized} is valid if and only if the quantum state is bipartite and pure, which can always be decomposed into a sum of product states factorized by an orthonormal bases $|\alpha _n\rangle_{\mathrm{A}}$ for subsystem A and an orthonormal basis $|\beta _n\rangle_{\mathrm{B}}$ for subsystem B
\begin{equation}
|\Psi \rangle _{\mathrm{AB}}=\sum_{n=1}^N{\sqrt{\lambda _n}|\alpha _n\rangle _{\mathrm{A}}\otimes |\beta _n\rangle _{\mathrm{B}}}
\label{equ:4.7}
\end{equation}
where the Schmidt coefficients $\sqrt{\lambda _n}$ are the non-negative numbers satisfying $\sum_{i=1}^N{\lambda _n}=1$. The orthonormal bases $|\alpha _n\rangle_{\mathrm{A}}$ and $|\beta _n\rangle_{\mathrm{B}}$ are known as the Schmidt bases of subsystems A and B respectively. The upper subscription of the summation $N=\min \left\{ N_{\mathrm{A}},N_{\mathrm{B}} \right\}$ is defined by the smaller one between the dimension of system A and the dimension of system B.

By comparing Eq.~\ref{equ:4.4} and Eq.~\ref{equ:4.7}, one can easily get that the separability criterion for a bipartite pure state is that only one of the Schmidt coefficients $\sqrt{\lambda _n}=1$ is 1 and the others are 0. The number of non-zero Schmidt coefficients is called the Schmidt number. Thus, a bipartite pure state is entangled if and only if its Schmidt number is larger than 1.

Schmidt decomposition is an application of singular value decomposition (SVD) in quantum mechanics. The practice and protocol of Schmidt decomposition are the same as SVD \cite{NielsenChuang}. One needs to trace out one of the subsystems and then diagonalize the reduced density matrix
\begin{subequations}
\begin{align}
\mathrm{Tr}_{\mathrm{A}}\left( |\Psi \rangle _{\mathrm{AB}}\langle \Psi |_{\mathrm{AB}} \right) =\rho _{\mathrm{B}}=\sum_{n=1}^N{\lambda _n|\beta _n\rangle _{\mathrm{B}}\langle \beta _n|_{\mathrm{B}}}
\\
\mathrm{Tr}_{\mathrm{B}}\left( |\Psi \rangle _{\mathrm{AB}}\langle \Psi |_{\mathrm{AB}} \right) =\rho _{\mathrm{A}}=\sum_{n=1}^N{\lambda _n|\alpha _n\rangle _{\mathrm{A}}\langle \alpha _n|_{\mathrm{A}}}
\end{align}
\label{equ:4.8}
\end{subequations}
The eigenvalues of $\rho _{\mathrm{B}}$ are actually the square of Schmidt coefficients $\lambda _n$ while the corresponding eigenvectors are the Schmidt Bases $|\beta _n\rangle_{\mathrm{B}}$. Use the same protocol to calculate reduced density matrix $\rho_{\mathrm{A}}$ and its eigensvalues $\lambda _n$ and eigenvectors $|\alpha _n\rangle_{\mathrm{A}}$. It is worth noticing that $\rho_{\mathrm{A}}$ and $\rho_{\mathrm{B}}$ share the same non-zero eigenspectrum $\lambda _n$. Many important properties of quantum systems such as expectation values of observables are completely determined by the eigenvalues of the density matrix, so for a bipartite pure state such properties of one subsystem will be identical to the other subsystem.\cite{NielsenChuang}

A bipartite pure state $|\Psi \rangle _{\mathrm{AB}}$ is separable if and only if its Schmidt number equals 1, so its reduced density matrix $\rho_{\mathrm{B}}$ can only have one non-zero eigenvalue, i.e. $\rho_{\mathrm{B}}$ is pure, which is the separable criterion for bipartite pure states. The purity $\gamma \left( \rho _{\mathrm{B}} \right) =\mathrm{Tr}\left( \rho _{\mathrm{B}}^{2} \right) =\sum_{n=1}^N{\lambda _{n}^{2}}$ of the reduced density matrix of a bipartite pure state can describe the separability and entanglement. $\gamma \left( \rho _{\mathrm{B}} \right)=1$ means separability and $\frac{1}{\min \left\{ N_{\mathrm{A}},N_{\mathrm{B}} \right\}}\leqslant \gamma \left( \rho _{\mathrm{B}} \right)<1$ means entanglement.

A more commonly used quantity for entanglement measurement in quantum information is the entanglement entropy, which is defined as the Von Neumann Entropy of the reduced density matrix
\begin{equation}
E_{\mathrm{S}}\left( \rho _{\mathrm{AB}} \right) =S\left( \rho _{\mathrm{B}} \right) =-\sum_{\lambda _n\ne 0}{\lambda _n\ln \lambda _n}
\label{equ:4.10}
\end{equation}
Entanglement entropy quantifies the entanglement of a bipartite pure state, i.e. larger $S\left( \rho _{\mathrm{B}} \right)$  indicates a more entangled state. A separable state has zero entanglement entropy $S\left( \rho _{\mathrm{B}} \right) =0$. A entangled state has positive entanglement entropy $S\left( \rho _{\mathrm{B}} \right) >0$. A full entangled state has the maximum entanglement entropy $S\left( \rho _{\mathrm{B}} \right) =\ln \left( \min \left\{ N_{\mathrm{A}},N_{\mathrm{B}} \right\} \right) $. Intuitively, the quantities to describe entanglement should not depend on the choice of the subsystem. Actually, $S\left( \rho _{\mathrm{B}} \right) =S\left( \rho _{\mathrm{A}} \right) $ and $\gamma\left( \rho _{\mathrm{B}} \right) =\gamma\left( \rho _{\mathrm{A}} \right) $, because entropy and purity are completely determined by the eigenvalues of the reduced density matrix, as stated above.

Here are some examples of the application of Schmidt decomposition and entanglement entropy. Consider a bipartite pure state $|\Psi _1\rangle _{\mathrm{AB}}=\frac{1}{2}\left( |00\rangle +|10\rangle -|01\rangle -|11\rangle \right) _{\mathrm{AB}}$. Using the protocol Eq.~\ref{equ:4.8} and Eq.~\ref{equ:4.10}, one can calculate $\rho _{1,\mathrm{B}}=\frac{1}{2}\left( |0\rangle \langle 0|+|1\rangle \langle 1|-|1\rangle \langle 0|-|0\rangle \langle 1| \right) _{\mathrm{B}}$, and $S\left( \rho _{1,\mathrm{B}} \right) =0$. Thus $|\Psi _1\rangle _{\mathrm{AB}}$ is a separable state. It can be further verified that $|\Psi _1\rangle _{\mathrm{AB}}=\frac{1}{\sqrt{2}}\left( |0\rangle +|1\rangle \right) _{\mathrm{A}}\otimes \frac{1}{\sqrt{2}}\left( |0\rangle -|1\rangle \right) _{\mathrm{B}}$, which indeed matches Eq.~\ref{equ:4.4}.

A second example is $|\Psi _2\rangle _{\mathrm{AB}}=\left( x|00\rangle +y|11\rangle \right) _{\mathrm{AB}}$, where $x$ and $y$ are the coefficients satisfying $\left| x \right|^2+\left| y \right|^2=1$. Then one can work out $\rho _{2,\mathrm{B}}=\left| x \right|^2|0\rangle _{\mathrm{B}}\langle 0|_{\mathrm{B}}+\left( 1-\left| x \right|^2 \right) |1\rangle _{\mathrm{B}}\langle 1|_{\mathrm{B}}$ and thus $S\left( \rho _{2,\mathrm{B}} \right) =-\left| x \right|^2\ln \left| x \right|^2-\left( 1-\left| x \right|^2 \right) \ln \left( 1-\left| x \right|^2 \right) $. When $x=\frac{1}{2}$, $S\left( \rho _{2,\mathrm{B}} \right) \approx 0.562$. When $x=\frac{1}{\sqrt{2}}$, $|\Psi _2\rangle _{\mathrm{AB}}$ is the Bell state and $S\left( \rho _{2,\mathrm{B}} \right) =\ln{2} \approx 0.693$. Both two states are entangled, but quantitatively the latter contains more entanglement than the former. Actually, all the four Bell states (Eq.~\ref{equ:qubit.2}) are maximally entangled as they have the maximum entanglement entropy $S\left( \rho _{\mathrm{B}} \right) =\ln{2}$.

\subsection{Entanglement Measures}
\citet{VedralPlenioKnight} proposed three conditions that any measures of entanglement $E\left(\rho\right)$ has to satisfy. Here we take bipartite entanglement as an example.

(1) Separable states have zero entanglement, i.e. $E\left( \sum_i{p_i\rho _{\mathrm{A}}^{i}\otimes \rho _{\mathrm{B}}^{i}} \right) =0$, because all entangled states can be purified to an ensemble of maximally entangled states, but separable states cannot.

(2) Local unitary operation, equivalent to local basis change, leaves entanglement invariant, i.e. $E\left( \rho _{\mathrm{AB}} \right) =E\left( U_{\mathrm{A}}\otimes U_{\mathrm{B}}\rho _{\mathrm{AB}}U_{\mathrm{A}}^{\dagger}\otimes U_{\mathrm{B}}^{\dagger} \right) $, because entanglement is basis-independent.

(3) Local operations and classical communication (LOCC) cannot increase entanglement, i.e. $E\left( \rho _{\mathrm{AB}} \right) \ge E\left( \sum_m{M_{\mathrm{A},m}\otimes M_{\mathrm{B},m}}\rho _{\mathrm{AB}}M_{\mathrm{A},m}^{\dagger}\otimes M_{\mathrm{B},m}^{\dagger} \right) $, where $\left\{ M_{\mathrm{A},m} \right\} $ and $\left\{ M_{\mathrm{B},m} \right\} $ are the local general measurement. The global general measurement $\left\{ M_{\mathrm{A},m}\otimes M_{\mathrm{B},m} \right\} $ (Eq.~\ref{equ:Measurement.DMGMDM}) makes $\left\{ M_{\mathrm{A},m} \right\} $ and $\left\{ M_{\mathrm{B},m} \right\} $ classically correlated (known as classical communication), and can only increase the classical-like correlation among the subsystems.

However, building general entanglement measures in this framework is difficult and computationally complex. So we are going to focus on the special cases. Apart from the entanglement entropy for pure bipartite states, we show some examples of non-standard entanglement measures for two-qubit entanglement. 

\paragraph{Concurrence}
Concurrence was first proposed by \citet{hill1997entanglement} based on the fidelity operator $R=\sqrt{\sqrt{\rho _{\mathrm{AB}}}\tilde{\rho}_{\mathrm{AB}}\sqrt{\rho _{\mathrm{AB}}}}$, where $\tilde{\rho}_{\mathrm{AB}}=\left( \sigma _{\mathrm{A}}^{\mathrm{y}}\sigma _{\mathrm{B}}^{\mathrm{y}} \right) \rho _{\mathrm{AB}}^{*}\left( \sigma _{\mathrm{A}}^{\mathrm{y}}\sigma _{\mathrm{B}}^{\mathrm{y}} \right) $ and $\rho _{\mathrm{AB}}^{*}$ is the complex conjugate of $\rho _{\mathrm{AB}}$ in $\sigma^{\mathrm{z}}$ basis. $R$ is Hermitian and has the eigenvalues $\lambda _{m}$ in the descending order. Concurrence is defined as
\begin{equation}
E_{\mathrm{C}}\left( \rho _{\mathrm{AB}} \right) =\max \left\{ 0,\lambda _1-\lambda _2-\lambda _3-\lambda _4 \right\} 
\label{equ:concurrence}
\end{equation}

\paragraph{Entanglement of Formation}
Entanglement of formation \cite{wootters1998entanglement} is the ensemble average of entanglement entropy:
\begin{equation}
E_{\mathrm{F}}\left( \rho _{\mathrm{AB}} \right) =\underset{\left\{ p_i,|\psi _i\rangle _{\mathrm{AB}} \right\}}{\min}\left\{ \sum_i{p_iE_{\mathrm{S}}\left( |\psi _i\rangle _{\mathrm{AB}} \right)} \right\} 
\label{equ:EntanglementofFormation}
\end{equation}
where the minimum runs over all the possible decomposition $\rho _{\mathrm{AB}}=\sum_i{p_i\left( |\psi _i\rangle \langle \psi _i| \right) _{\mathrm{AB}}}$ and $E_{\mathrm{S}}\left( |\psi _i\rangle _{\mathrm{AB}} \right) $ is the entanglement entropy (Eq.~\ref{equ:4.10}). For the two-qubit case, it has the closed-form expression
\begin{equation}
E_{\mathrm{F}}\left( \rho _{\mathrm{AB}} \right) =H\left( \left\{ \frac{1+\sqrt{1-E_{\mathrm{C}}\left( \rho _{\mathrm{AB}} \right) ^2}}{2},\frac{1-\sqrt{1-E_{\mathrm{C}}\left( \rho _{\mathrm{AB}} \right) ^2}}{2} \right\} \right) 
\label{equ:EntanglementofFormationtwoqubit}
\end{equation}
where $E_{\mathrm{C}}\left( \rho _{\mathrm{AB}} \right) $ is the concurrence (Eq.~\ref{equ:concurrence}) and $H\left( \left\{ p_1,p_2 \right\} \right) $ is the Shannon entropy (Eq.~\ref{equ:Entropy.Shannon}) for binomial distribution.

\paragraph{Relative Entropy of Entanglement}
Relative entropy of entanglement \cite{Vedral2002} is a distance measure between the entangled state $\rho _{\mathrm{AB}}$ and the separable set $\mathcal{S} _1$.
\begin{equation}
E_{\mathrm{R}}\left( \rho _{\mathrm{AB}} \right) =\underset{\rho _{\mathrm{s},\mathrm{AB}}\in \mathcal{S} _1}{\min}\left\{ \mathrm{Tr}\left( \rho _{\mathrm{AB}}\ln \frac{\rho _{\mathrm{AB}}}{\rho _{\mathrm{s},\mathrm{AB}}} \right) \right\} 
\label{equ:RelativeEntropyofEntanglement}
\end{equation}
In the absence of a closed-form expression, \citet{VedralPlenioKnight} discussed the properties of $E_{\mathrm{R}}\left( \rho _{\mathrm{AB}} \right)$ for a two-qubit system.

\paragraph{Negativity}

Negativity is a computationally feasible entanglement measure \cite{vidal2002computable}, originating from the violation of "positive semi-definite" of the reduced density matrix of an entangled state.
\begin{equation}
E_{\mathrm{N}}\left( \rho _{\mathrm{AB}} \right) =\frac{\left\| \rho _{\mathrm{B}} \right\| -1}{2}
\label{equ:Negativity}
\end{equation}
where $\rho _{\mathrm{B}}=\mathrm{Tr}_{\mathrm{A}}\left( \rho _{\mathrm{AB}} \right) $ is the reduced density matrix and $\left\| \rho _{\mathrm{B}} \right\| =\mathrm{Tr}\sqrt{\rho _{\mathrm{B}}^{\dagger}\rho _{\mathrm{B}}}$ is the trace norm.

\subsection{Entanglement Witness}
In general, it is NP-hard to find the sufficient and necessary criterion of entanglement qualitatively \cite{gurvits2004classical}. Moving a step backward, there are some sufficient but not necessary conditions for entanglement, regardless of the dimension, known as entanglement witnesses \cite{friis2019entanglement, guhne2009entanglement, terhal2000bell}. An entanglement witness is a Hermitian operator $\mathcal{W\left(\rho\right)}$ that can sufficiently distinguish certain entangled states $\rho$ from all the separable states. Specifically, entanglement witnesses are defined as the Hermitian operators that obey the inequality $\mathrm{Tr}\left( \mathcal{W} \left( \rho \right) \rho \right) \ge 0 $ for all the separable states. Thus a negative expectation value $\mathrm{Tr}\left( \mathcal{W} \left( \rho \right) \rho \right)<0 $ means that $\rho$ must be entangled. $\mathcal{W\left(\rho\right)}$ may be linear or nonlinear depending on $\rho$.

The convexity of the set of all separable states ensures the existence of the witness operator $\mathcal{W\left(\rho\right)}$ and the validation of an entanglement witness \cite{horodecki2009quantum}. The Hahn–Banach theorem guarantees that there exists a hyperplane for any entangled state $\rho$ that distinguishes this state from the separable set \cite{friis2019entanglement}, as shown in Figure.~\ref{fig:EW}, where the hyperplanes are represented by the green dotted lines. These hyperplanes correspond to observables $\mathcal{W} \left( \rho \right)$, represented by $\mathrm{Tr}\left( \mathcal{W} \left( \rho \right) \rho \right) = 0 $. If there is one separable state that makes $\mathrm{Tr}\left( \mathcal{W} \left( \rho \right) \rho \right) = 0 $, the hyperplane is at a tangent to the convexity set of separable states. Such tangent hyperplanes are usually thought to be efficient entanglement witnesses that reach their tight bound.

It is worth noticing that an entanglement witness is just a sufficient but not necessary condition that qualitatively describes entanglement \cite{Peres1996}. $\mathrm{Tr}\left( \mathcal{W} \left( \rho \right) \rho \right)<0 $ means that $\rho$ must be entangled, while the inverse and the converse may not be true. As shown in Figure.~\ref{fig:EW}, some entangled state states satisfy $\mathrm{Tr}\left( \mathcal{W} \left( \rho \right) \rho \right) \ge 0$. Thus, $\mathrm{Tr}\left( \mathcal{W} \left( \rho \right) \rho \right) \ge 0$ does not imply that $\rho$ is separable. Entanglement witnesses can only confirm entangled states but cannot exclude that a state is entangled. For the same reason, the expectation value $\mathrm{Tr}\left( \mathcal{W} \left( \rho \right) \rho \right)$ cannot quantify entanglement; a more negative value of $\mathrm{Tr}\left( \mathcal{W} \left( \rho \right) \rho \right)$ does not suggest $\rho$ is more entangled. 

Bell’s theorem \cite{bell1964einstein} is a famous example of entanglement witness. There are many different formalisms of Bell’s inequality. Here we focus on the CHSH (Clauser-Horne-Shimony-Holt) inequality \cite{collins2004relevant}. Consider a total system $\rho_{\mathrm{AB}}$ consisting of two half-spins. If $\rho_{\mathrm{AB}}$ is separable, the CHSH inequality holds
\begin{equation}
\mathrm{Tr}\left( \rho \mathcal{W} _{\mathrm{CHSH}} \right) =2+\left< \sigma _{\mathrm{A}}^{\alpha _2}\sigma _{\mathrm{B}}^{\beta _2} \right> -\left< \sigma _{\mathrm{A}}^{\alpha _1}\sigma _{\mathrm{B}}^{\beta _1} \right> -\left< \sigma _{\mathrm{A}}^{\alpha _2}\sigma _{\mathrm{B}}^{\beta _1} \right> -\left< \sigma _{\mathrm{A}}^{\alpha _1}\sigma _{\mathrm{B}}^{\beta _2} \right> \geqslant 0
\label{equ:CHSH}
\end{equation}
where $\sigma ^{\alpha}=\vec{\alpha}\cdot \vec{\sigma}$ and $\left< \sigma _{\mathrm{A}}^{\alpha}\sigma _{\mathrm{B}}^{\beta} \right> =\mathrm{Tr}\left( \rho _{\mathrm{AB}}\sigma _{\mathrm{A}}^{\alpha}\sigma _{\mathrm{B}}^{\beta} \right) $.
Although separable states always satisfy the CHSH inequality, entangled states may disobey Bell’s theorem. In this sense, the CHSH inequality (and other Bell’s inequalities) can work as entanglement witnesses, because a quantum state not satisfying Eq.~\ref{equ:CHSH} must be entangled. We can test CHSH inequality by taking the unit vector $\alpha _1=\left( 0,0,1 \right) $, $\alpha _2=\left( 1,0,0 \right) $, $\beta _1=\left( -\frac{1}{\sqrt{2}},0,-\frac{1}{\sqrt{2}} \right) $, $\beta _2=\left( \frac{1}{\sqrt{2}},0,-\frac{1}{\sqrt{2}} \right) $, and the singlet state $|\psi \rangle =\frac{1}{\sqrt{2}}\left( |01\rangle -|10\rangle \right) _{\mathrm{AB}}$. We then work out $\langle \psi |\mathcal{W} _{\mathrm{CHSH}}|\psi \rangle  =2\left(1-\sqrt{2}\right)<0$, which proves that the singlet state is entangled. This agrees with the entanglement entropy for the Bell state. 

The choice of entanglement witness operator $\mathrm{Tr}\left( \mathcal{W} \left( \rho \right) \rho \right)$ is not unique, because there are infinite witness hyperplanes in the Liouville space. For example, there is an arbitrariness in choosing the direction of the unit vector $\vec{\alpha}$ and $\vec{\beta}$ in Eq.~\ref{equ:CHSH}. Apart from Bell's theorem, variance and quantum Fisher information have also been considered to build entanglement witnesses. Among so many entanglement witnesses, their performance varies based on the system of interest. Specifically, entanglement witnesses with negative expectation values regarding the system of interest are good entanglement witnesses. Figure.~\ref{fig:EW} shows two examples, $\mathcal{W}_{\mathrm{g}}$ and $\mathcal{W}_{\mathrm{b}}$, where $\mathcal{W}_{\mathrm{g}}$ is a good entanglement witness for the system of interest but $\mathcal{W}_{\mathrm{b}}$ is not a suitable witness for the underlying state.

The example of the CHSH inequality also shows that some witnesses are not good when studying the singlet state. If we take $\alpha _1=\left( 0,0,1 \right) $, $\alpha _2=\left( 1,0,0 \right) $,  $\beta _1=\beta _2= \left( 0,0,1\right) $ in Eq.~\ref{equ:CHSH}, the singlet state $|\psi \rangle =\frac{1}{\sqrt{2}}\left( |01\rangle -|10\rangle \right) _{\mathrm{AB}}$ gives the expectation value of $\langle \psi |\mathcal{W} _{\mathrm{CHSH}}|\psi \rangle =2-\sqrt{2}\ge 0$. The challenge hence lies in the construction of useful entanglement witnesses, tailored for the systems and properties of interest produced in an experiment. In the Bell test, the correlation between different spin projections is measured to get a powerful entanglement witness. 

Entanglement witnesses also offer a feasible way to advise experimentists on how to prove the existence of entanglement in a quantum system, because we only need to detect the expectation value of a suitable observable. Entanglement witnesses are usually treated as one of the most important practical entanglement certification techniques \cite{friis2019entanglement}. For example, Aspect's experiment is a Bell test that proves the violation of Bell inequalities by measuring the correlation between entangled photons \cite{aspect1976proposed}. It was recently been shown that quantum Fisher information of certain states of a molecular aggregate or polariton system can be measured via linear absorption spectrum \cite{sifain2021toward}, and an entanglement witness operator $\mathcal{W}_{\mathrm{QFI}}\left(\rho\right)$ can be built by quantum Fisher information. 

Take polariton chemistry as an example. It has been shown that in the normal regime (the typical light-matter coupling strength), $\mathcal{W}_{\mathrm{QFI}}\left(\rho\right)$ works as a good multi-molecular entanglement witness at room temperature and can be measured by bounded mode absorption. This is rigorous proof of the existence of long-lived multi-molecular entanglement because of intramolecular interaction indirectly induced by cavity QED, despite the dephasing due to the system-bath interaction at room temperature \cite{wu2023molecular}.

\section{Correlation and Coherence}
In classical statistics, covariance describes the correlation between two random variables $X$ and $Y$: $\mathrm{cov}\left[ X,Y \right] =E\left[ XY \right] -E\left[ X \right] E\left[ Y \right]$. In quantum mechanics, to study the correlation between two quantities or operators, the correlation function can be similarly written as $\mathrm{Tr}\left( \rho O_1O_2 \right) -\mathrm{Tr}\left( \rho O_1 \right) \mathrm{Tr}\left( \rho O_2 \right) $. It is convenient to assume $O_1$ and $O_2$ have zero expectation value. Otherwise, we may redefine the operators by shifting the operators with a constant as $O^{\prime}=O-\mathrm{Tr}\left( \rho O \right) $. Typically, the two-point correlation function is defined as
\begin{equation}
\left< O_1 O_2 \right> =\mathrm{Tr}\left( \rho O_1 O_2 \right) 
\label{equ:correlator}
\end{equation}
where the operators do not have to be observable.

The quantification of correlation is well-defined and has been widely studied in physics and chemistry. For example, in molecular spectroscopy, it is predicted using auto-correlation of the dipole operator \cite{bloembergen1982nonlinear,shen1984principles} as $\left< \mu \left( t \right) \mu \left( 0 \right) \right> $. In many-body physics and field theory, the two-point correlator between creation and annihilation operators in the time-order $\left< Ta_k\left( t \right) a_{k^{\prime}}^{\dagger}\left( t^{\prime} \right) \right> $ is the time-ordered Green's function and propagator \cite{peskin2018introduction}. In quantum optics, correlation functions are used to characterize the statistical and coherence properties of an electromagnetic field $\mathscr{E} \left( \vec{r},t \right)$, known as the degree of coherence \cite{zernike1938concept}  $g^{\left( 1 \right)}=\frac{\left< \mathscr{E} ^*\left( \vec{r}_1,t_1 \right) \mathscr{E} \left( \vec{r}_2,t_2 \right) \right>}{\sqrt{\left< \left| \mathscr{E} \left( \vec{r}_1,t_1 \right) \right|^2 \right> \left< \left| \mathscr{E} \left( \vec{r}_1,t_1 \right) \right|^2 \right>}}$, where the phase is encoded in the complex field $\mathscr{E}$. Similarly, the $n$-point correlation function can be generalized, for example, to study the higher-order Dyson series in non-linear spectroscopy or the degree of second-order coherence \cite{zernike1938concept} $g^{\left( 2 \right)}=\frac{\left< \mathscr{E} ^*\left( \vec{r}_1,t_1 \right) \mathscr{E} ^*\left( \vec{r}_2,t_2 \right) \mathscr{E} \left( \vec{r}_1,t_1 \right) \mathscr{E} \left( \vec{r}_2,t_2 \right) \right>}{\left< \left| \mathscr{E} \left( \vec{r}_1,t_1 \right) \right|^2 \right> \left< \left| \mathscr{E} \left( \vec{r}_1,t_1 \right) \right|^2 \right>}$. 

In this paper, we are more interested in the correlation between two sub-systems, to clarify the difference between quantum correlation and classical correlation, and their relation to the entanglement between two sub-systems. In this case, we take $O_{\mathrm{A}}$ and $O_{\mathrm{B}}$ as two local operators on two sub-systems in Eq.~\ref{equ:correlator}. The concept of quantum coherence is also clarified because when calculating the correlation function, the phase-induced interference shows up.

\subsection{Quantum Correlation}
Two subsystems are entangled to each other implying a quantum correlation between them. Quantum entanglement is a special form of quantum correlation without classical analogy. To further clarify this statement, a quantum state composed of an uncorrelated pair of sub-systems is shown below:
\begin{equation}
    \rho _{\mathrm{AB}}=\rho _{\mathrm{A}}\otimes \rho _{\mathrm{B}}
\label{equ:4.11}
\end{equation}
If one does a local measurement on subsystem A by a local operator $O_{\mathrm{A}}$, the expectation value $\left< O_{\mathrm{A}} \right> =\mathrm{Tr}\left( O_{\mathrm{A}}\rho _{\mathrm{AB}} \right) =\mathrm{Tr}\left( O_{\mathrm{A}}\rho _{\mathrm{A}} \right)$ is completely determined by the information about subsystem A and no information will be gained for B from this measurement. No mutual information is lost during the measurement. Similarly, the two-point correlator of an uncorrelated pair is $\left< O_{\mathrm{A}}O_{\mathrm{B}} \right> =\left< O_{\mathrm{A}} \right> \left< O_{\mathrm{B}} \right> $, which is simply the multiplication of two expectation value with no cross-term. Thus, we can reconstruct the uncorrelated quantum state $\rho _{\mathrm{AB}}$ by doing quantum tomography independently on subsystems A and B. Eq.~\ref{equ:4.11} defines the uncorrelated quantum states. A classical analogy to uncorrelated quantum systems is independence in classical probability theory, where the joint probability of a pair of random variables $\left( X, Y \right) $  can be written as the product of their probabilities: $P_{XY}\left( x,y \right) =P_X\left( x \right) P_Y\left( y \right) $.

\begin{figure}[!h]
    \centering\includegraphics[width=0.9\textwidth]{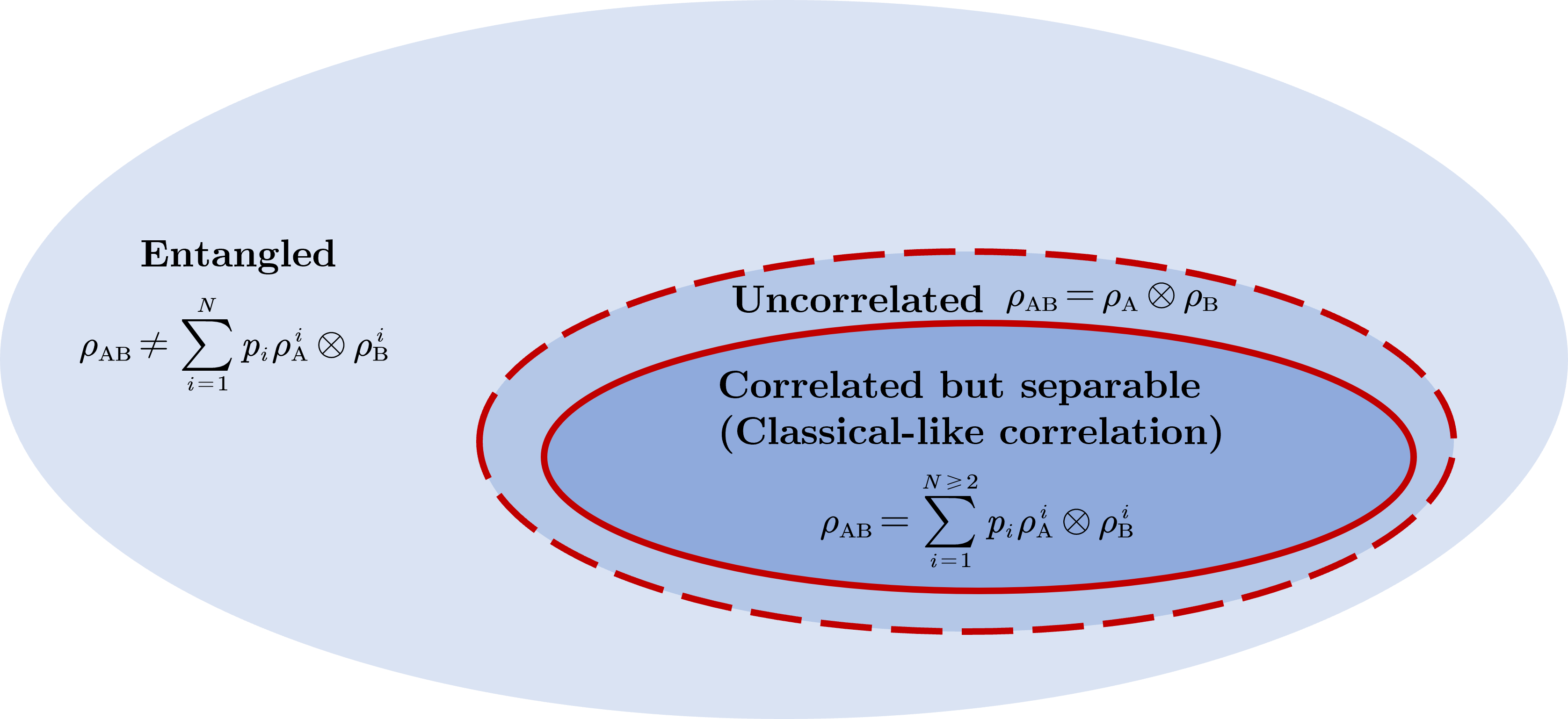}
    \caption{A set of all the bipartite states (Liouville space) as an illustration for entanglement, correlation. The red solid curve is the boundary between correlated states and uncorrelated states. The red dashed curve is the boundary between entangled states and separable states.}
    \label{fig:Liouville}
\end{figure}

A state with quantum correlation between two subsystems is then defined as a quantum state that is not uncorrelated:
\begin{equation}
    \rho _{\mathrm{AB}}\ne\rho _{\mathrm{A}}\otimes \rho _{\mathrm{B}}
\label{equ:4.119}
\end{equation}
The distinction between correlated and uncorrelated systems is demonstrated in Figure.~\ref{fig:Liouville} by the red solid curve. 

As stated before, quantum entanglement is a special form of quantum correlation. Similarly, by comparing separable states (Eq.~\ref{equ:4.5}) and uncorrelated states (Eq.~\ref{equ:4.11}), one can find that the uncorrelated state is a special form of the separable state. Uncorrelated states form the convex hull of the separate set. Therefore, there exists a set of density matrices that is correlated but separable (not entangled), known as classical-like correlation, as shown in Figure.~\ref{fig:Liouville}:
\begin{equation}
    \rho _{\mathrm{AB}}=\sum_{i=1}^{n\geqslant 2}{p_i\rho _{\mathrm{A}}^{i}\otimes \rho _{\mathrm{B}}^{i}}
\label{equ:4.12}
\end{equation}
where $p_i>0$ and $\sum_i{p_i}=1$ with at least 2 non-zero terms. The classical analogy in probability theory is  $P_{XY}\left( x,y \right) =\sum_{i=1}^{n\geqslant 2}{p_iP_{X}^{i}\left( x \right) P_{Y}^{i}\left( y \right)}$. An example of a correlated but not entangled state is 
\begin{equation}
    \rho _{\mathrm{AB}}^{\left( 1 \right)}=\frac{1}{2}\left( |00\rangle \langle 00|+|11\rangle \langle 11| \right) _{\mathrm{AB}}
\label{equ:4.13}
\end{equation}
if we do a one-time measurement on subsystem A by the local operator $\sigma _{\mathrm{A}}^{\mathrm{z}}$ and get $|0\rangle _{\mathrm{A}}$, we can predict that the following measurement by the local operator $\sigma _{\mathrm{B}}^{\mathrm{z}}$ will result in $|0\rangle _{\mathrm{B}}$, and vice versa. Therefore, $\rho _{\mathrm{AB}}^{\left( 1 \right)}$ shows a strong correlation between subsystems A and B, despite it being separable. This correlation is classical-like.

On the contrary, a Bell state $|\Phi _+\rangle _{\mathrm{AB}}=\frac{1}{\sqrt{2}}\left( |00\rangle +|11\rangle \right) _{\mathrm{AB}}$, or $\rho _{\mathrm{AB}}^{\left( 2 \right)}=|\Phi _+\rangle _{\mathrm{AB}}\langle \Phi _+|_{\mathrm{AB}}$, is entangled. However, $\rho _{\mathrm{AB}}^{\left( 2 \right)}$ contains some similar inter-system correlation to $\rho _{\mathrm{AB}}^{\left( 1 \right)}$, because if we do a local projective measurement $\sigma _{\mathrm{A}}^{\mathrm{z}}$ first, then followed by $\sigma _{\mathrm{B}}^{\mathrm{z}}$, we will get the same results as for $\rho _{\mathrm{AB}}^{\left( 1 \right)}$. Thus, entangled states and classical-like correlated states share the same local measurement outcome and the same correlation function, $ \left< \sigma _{\mathrm{A}}^{\mathrm{z}}\sigma _{\mathrm{B}}^{\mathrm{z}} \right> ^{\left( 1 \right)}=\left< \sigma _{\mathrm{A}}^{\mathrm{z}}\sigma _{\mathrm{B}}^{\mathrm{z}} \right> ^{\left( 2 \right)}=1$, which means correlation is not solid evidence of entanglement:

To better understand this, we write the Bell state in density matrix framework $\rho _{\mathrm{AB}}^{\left( 2 \right)}=|\Phi _+\rangle _{\mathrm{AB}}\langle \Phi _+|_{\mathrm{AB}}$ and expand it:
\begin{equation}
    \rho _{\mathrm{AB}}^{\left( 2 \right)}=\frac{1}{2}\left( |00\rangle \langle 00|+|11\rangle \langle 11|+|00\rangle \langle 11|+|11\rangle \langle 00| \right) _{\mathrm{AB}}
\label{equ:4.14}
\end{equation}
It is now clear that $\rho _{\mathrm{AB}}^{\left( 2 \right)}$ has non-zero quantum coherence in the chosen bases, while $\rho _{\mathrm{AB}}^{\left( 1 \right)}$ doesn't. So, it is the quantum coherence that induces the non-classical entanglement. When performing local projective measurements on one subsystem, we only keep the reduced density and lose coherence, which is actually a dephasing process. $\rho _{\mathrm{AB}}^{\left( 2 \right)}$ will dephase to $\rho _{\mathrm{AB}}^{\left( 1 \right)}$ when it strongly interacts with the measurement instrument. We can further check that $\rho _{\mathrm{AB}}^{\left( 1 \right)}$ and $\rho _{\mathrm{AB}}^{\left( 2 \right)}$ share the same reduced density matrix: 
\begin{equation}
\mathrm{Tr}_{\mathrm{B}}\left( \rho _{\mathrm{AB}}^{\left( 1 \right)} \right) =\mathrm{Tr}_{\mathrm{B}}\left( \rho _{\mathrm{AB}}^{\left( 2 \right)} \right) =\frac{1}{2}\left( |0\rangle \langle 0|+|1\rangle \langle 1| \right) _{\mathrm{A}}
\label{equ:4.15}
\end{equation}

Although quantum correlation is a more general concept than quantum entanglement, they are related. Correlation functions can be used as entanglement witnesses that sufficiently distinguish some entangled states, and the correlation functions are easy to measure and quantify. As an example discussed before, Bell's inequality (Eq.~\ref{equ:CHSH}) offers some constraint of the two-point correlation functions for the separable states, including those classical-like correlated states, but the entangled states are not bounded by it. In the Bell test, the quantum correlation is revealed by noting that the singlet spin state is unchanged by observing it in any reference frame (e.g. the observer can rotate the axes and still measure the same spin-pairing). Bell realized that the extra quantum correlation can often (but not always) be detected by comparing two measurements performed in different reference frames \cite{Ballentine}.

\subsection{Quantum Coherence}
The previous example shows that the entangled state is not only correlated but also a violation of locality and realism. The distinction between classical-like correlation and quantum entanglement can be understood by quantum coherence. The concept of coherence originates from wave interference, where the phase difference leads to constructive or destructive interference. In quantum mechanics, a pure state is just a wavefunction. As shown in Eq.~\ref{equ:qubit.1}, $\langle 1|O|0\rangle $ shows the effect of interference and $c_0c_{1}^{*}$ quantifies the magnitude of interference, known as quantum coherence. 

More rigorously, the concept of quantum coherence needs to be discussed in the density matrix formalism. Quantum coherence is signaled by the off-diagonal terms of a density matrix in a particular basis. When studying a density matrix in this fixed basis, the diagonal term is considered a series of classical-like probabilities, while the off-diagonal terms emphasize the quantum properties of the state. Thus, a diagonal density matrix is considered classical-like in this fixed basis. When doing PVM, we project the quantum state $\rho$ to this fixed basis by keeping only the diagonal terms and dropping all the coherence, resulting in an incoherent and classical-like quantum state $\rho ^{\prime}=\sum_i{P_m\rho P_m}$.

It is important to point out that coherence varies in different bases of the same quantum state. So clarifying the basis is the first step to study coherence. 

To quantify quantum coherence in a fixed basis, we follow the rule of monotonicity under incoherent operations (quantum operations that map the set of incoherent states onto itself). \citet{baumgratz2014quantifying} proposed the $l_1$-norm coherence
\begin{equation}
C_{l_1}\left( \rho \right) =\sum_{m_1,m_2}{\left| \langle m_1|\left( \rho -\rho ^{\prime} \right) |m_2\rangle \right|}=\sum_{m_1\ne m_2}{\left| \rho _{m_1,m_2} \right|}
\label{equ:4.l1}
\end{equation}
which is the sum of all the $l_1$-norm of the off-diagonal terms. The relative entropy coherence is defined as
\begin{equation}
C_{\mathrm{r}.\mathrm{e}.}\left( \rho \right) =\mathrm{Tr}\left( \rho \ln \left( \frac{\rho}{\rho ^{\prime}} \right) \right) =S\left( \rho ^{\prime} \right) -S\left( \rho \right) 
\label{equ:4.r.e.}
\end{equation}
which is just the difference between the Von Neumann entropy of $\rho ^{\prime}$ and $\rho$, where the positivity is guaranteed by Eq.~\ref{equ:Entropy.PVM}.

Although relevant, quantum coherence is very different from quantum correlation and quantum entanglement. A feature of either quantum correlation (uncorrelatedness) or quantum entanglement (separability) is basis independent, unlike quantum coherence. Intuitively, the correlation or entanglement between the two subsystems should not depend on the basis chosen to describe the composite system.

\section{Conclusion}
In this article, we reviewed some basic concepts in quantum information by providing rigorous mathematical definitions and physical pictures. We attempted to frame the paper especially for researchers in chemistry and other fields outside of quantum physics. In particular, we emphasized how the basic tools are needed to deal with the generalization of classical probability theory in classic information theory to density matrix theory, which is central to quantum information.

With the definition of mixed states, quantum measurements, and the von Neumann entropy, entanglement can be rigorously discussed. Although chemistry is dominated by quantum mechanics, many chemical systems are classical-like and can be well described in the classical context. Entanglement is of great interest because it has no classical analogy, which implies that it can be exploited in ways not possible for classical systems. Entanglement is defined as a state that cannot be written as a convex combination of the tensor product states of the subsystems. For a bipartite pure state, entanglement entropy (Eq.~\ref{equ:4.10}) quantifies entanglement. While small model systems can be well studied, chemistry will likely shift focus to much larger quantum systems. A key point to realize is that there is no simple way to describe entanglement for arbitrary numbers of particles and arbitrary dimensions. A sufficient approach to qualitatively describe multipartite entanglement involves entanglement witnesses, such as Bell's inequality \cite{aspect1976proposed}, variance \cite{friis2019entanglement}, or quantum Fisher information \cite{wu2023molecular,sifain2021toward}. These kinds of techniques and developments thereof will be of interest for classifying chemical entanglement.

It is sometimes easy to get confused between quantum entanglement, quantum correlation, and quantum coherence, which, though related, are fundamentally different. Quantum entanglement is a special form of quantum correlation, but correlation does not always mean entanglement. The disentangled correlation is known as the classical-like correlation. Quantum correlation can be easily quantified by correlation function both theoretically and experimentally, but it is usually an NP-hard problem (i.e. it cannot be solved in polynomial time with respect to the size of the system) to describe entanglement. Quantum entanglement is the quantum correlation with no classical analogy, where the non-trivial quantum effect is guaranteed by quantum coherence in a fixed basis. However, quantum coherence does not mean quantum entanglement, because entanglement and correlation are basis-independent, but coherence varies on the choice of basis. 

Quantum correlation can be used to build entanglement witnesses. For example, the Bell test aims to measure the correlation between spins in different directions to sufficiently find entanglement. Linear absorption spectroscopy can measure the dipole auto-correlator to witness entanglement. As anticipated above, quantum entanglement defies explanation by classical-like correlation. Moreover, we showed how certain types of quantum correlations come hand-in-hand with their possible classical analogies, and often the classical correlations dominate, for example, the adiabatic dynamics and Marokov dynamics. Physically, this is the same idea as exemplified by interference, as seen in the double-slit experiment. Interference is insufficient evidence for uniquely quantum correlations, even though it evidences wave-like properties. 

There are many examples of classical correlations exhibited by chemical systems. These include synchronization produced by feedback loops, oscillating reactions, or any switched system under kinetic control. Although these are interesting phenomena, they are not directly relevant to the non-trivial features of quantum information. To build the non-trivial features of entanglement between molecules, inter-molecular interaction is required, and entanglement grows at a rate proportional to the interaction \cite{li2023speeding}. However, in chemistry, due to the noisy environment (i.e., the system-bath interaction), large dephasing destroys entanglement, rendering the system classical-like. Such chemical systems show mostly classical behavior and lack the long-lived quantum entanglement needed in quantum information applications. So we need to understand entanglement to engineer chemical systems for quantum information. To build entanglement and maintain it for a longer time, larger inter-molecular interactions that can compete with dephasing is a way forward, for example, in molecular aggregates \cite{hestand2018expanded} or cavity QED \cite{frisk2019ultrastrong, forn2019ultrastrong, mandal2023theoretical}. Entanglement witnesses such as quantum Fisher information\cite{wu2023molecular} could be useful to sufficiently prove the existence of entanglement.

We hope this detailed mini-review serves as a helpful primer to provide rigorous background on quantum information. That, in turn, will serve as a platform for discovering new examples of, and tools to study, quantum information in the molecular domain.

\begin{acknowledgement}
This material is based upon work supported by the National Science Foundation under Grant No. 2211326. The authors thank Ava N. Hejazi and Alfy Benny for the valuable discussion and review.

\end{acknowledgement}


\bibliography{achemso-demo}

\end{document}